\documentclass[twocolumn,longbibliography,showpacs,preprintnumbers,amsmath,amssymb]{revtex4-1}

\usepackage{dcolumn}% Align table columns on decimal point
\usepackage{bm}% bold math

\usepackage{color}

\usepackage{graphicx}
\usepackage{natbib}
\usepackage{subfigure}
\usepackage{cases}
\usepackage{epstopdf}
\newcommand{\beq}{\begin{equation}}
\newcommand{\eeq}{\end{equation}}

\begin{document}
\def\bfB{\mbox{\bf B}}
\def\bfQ{\mbox{\bf Q}}
\def\bfD{\mbox{\bf D}}
\def\etal{\mbox{\it et al}}
%
%\title{From weak MHD turbulence to 2D turbulence}
\title{On the 2D behavior of 3D MHD with a strong guiding field}
\author{Alexandros Alexakis}

%\address
\affiliation{Laboratoire de Physique Statistique de l'Ecole Normale
Sup\'erieure, UMR CNRS 8550, 24 Rue Lhomond, 75006 Paris Cedex 05, France.}

\date{\today}

\begin{abstract}
The Magneto-hydrodynamic (MHD) equations in the presence of a guiding magnetic field 
are investigated by means of direct numerical simulations.
The basis of the investigation consists of 9 runs forced at the small scales. 
The results demonstrate that for a large enough  uniform magnetic field the large scale flow behaves as a two dimensional 
(non-MHD) fluid exhibiting an inverse cascade of energy in the direction perpendicular to the magnetic field,
while the small scales behave like a three dimensional MHD-fluid cascading the energy forwards.
The amplitude of the inverse cascade is sensitive to the magnetic field amplitude, the domain size, the forcing mechanism,
and the forcing scale.
All these dependencies are demonstrated by the varying parameters of simulations.
Furthermore, in the case that the system is forced anisotropically in the small parallel scales an inverse cascade in the 
parallel direction is observed that is feeding the 2D modes $k_\|=0$.
\end{abstract} 
%
%
%\pacs{05.45.-a, 91.25.Cw}
%

\maketitle

%%%%%%%%%%%%%%%%%%%%%%%%%%%%%%%%%%%%%%%%%%%%%%%%%%%%%%%%%%%%%%%%%%%%%%%%%%%%%%%%%%%%%%%%%%%%%%%%%%%%%%%%%%%%%%%%%%%%%%%%%%%%%%%%
%%%%%%%%%%%%%%%%%%%%%%%%%%%%%%%%%%%%%%%%%%%%%%%%%%%%%%%%%%%%%%%%%%%%%%%%%%%%%%%%%%%%%%%%%%%%%%%%%%%%%%%%%%%%%%%%%%%%%%%%%%%%%%%%
%%%%%%%%%%%%%%%%%%%%%%%%%%%%%%%%%%%%%%%%%%%%%%%%%%%%%%%%%%%%%%%%%%%%%%%%%%%%%%%%%%%%%%%%%%%%%%%%%%%%%%%%%%%%%%%%%%%%%%%%%%%%%%%%
%%%%%%%%%%%%%%%%%%%%%%%%%%%%%%%%%%%%%%%%%%%%%%%%%%%%%%%%%%%%%%%%%%%%%%%%%%%%%%%%%%%%%%%%%%%%%%%%%%%%%%%%%%%%%%%%%%%%%%%%%%%%%%%%
\section{Introduction}
%%%%%%%%%%%%%%%%%%%%%%%%%%%%%%%%%%%%%%%%%%%%%%%%%%%%%%%%%%%%%%%%%%%%%%%%%%%%%%%%%%%%%%%%%%%%%%%%%%%%%%%%%%%%%%%%%%%%%%%%%%%%%%%%
%%%%%%%%%%%%%%%%%%%%%%%%%%%%%%%%%%%%%%%%%%%%%%%%%%%%%%%%%%%%%%%%%%%%%%%%%%%%%%%%%%%%%%%%%%%%%%%%%%%%%%%%%%%%%%%%%%%%%%%%%%%%%%%%
%%%%%%%%%%%%%%%%%%%%%%%%%%%%%%%%%%%%%%%%%%%%%%%%%%%%%%%%%%%%%%%%%%%%%%%%%%%%%%%%%%%%%%%%%%%%%%%%%%%%%%%%%%%%%%%%%%%%%%%%%%%%%%%%
%%%%%%%%%%%%%%%%%%%%%%%%%%%%%%%%%%%%%%%%%%%%%%%%%%%%%%%%%%%%%%%%%%%%%%%%%%%%%%%%%%%%%%%%%%%%%%%%%%%%%%%%%%%%%%%%%%%%%%%%%%%%%%%%
The existence of magnetic fields is known in many astrophysical
objects, such as the interstellar medium, galaxies,
accretion disks, star and planet interiors and the solar wind  
\cite{Zeldo1990}. In most of these systems, the magnetic fields are
strong enough to play a significant dynamical role. 
The kinetic and magnetic Reynolds numbers involved in these astrophysical
bodies are large enough so that the flows exhibit
a turbulent behavior with a large continuous range of excited
scales, from the largest where energy is injected, toward the
finest where energy is dissipated. 
In many cases, strong large-scale magnetic fields are present that induce dynamic anisotropy in the small scales. 
Direct numerical simulations that examine in detail both large and small scale turbulent processes in astrophysical plasmas
are very difficult to achieve, and only modest scale separation can be reached even with today's super-computers. 
One way around this difficulty is to model the effect of the large scale field by a uniform magnetic field $B_0$, and
thus study the small scales separately. 

In the presence of a strong uniform magnetic field the evolution of the turbulent 
fluctuating fields can be treated within the framework of weak turbulence theory (WTT). In this approach the nonlinearities
are treated perturbatively, resulting in a slowly varying amplitude of the linear wave modes
that in this case are the Alfven waves supported by the uniform magnetic field \cite{Galtier2000,Nazarenko_book2011}.
However, the limit $B_0\to \infty$ is non-trivial because different limiting procedures can lead to different
results.
Thus in order for the results of weak turbulence theory \cite{Galtier2000} to hold a domain of size $L$ 
(in direction of the field) sufficiently large needs to be considered, so that modes with small wave numbers in the direction 
of the field that satisfy the ``quasi-resonance" conditions are present. 
In particular WTT is valid when the following condition is met:
\beq
\frac{1}{\sqrt{ k_\|  L_\|} } \ll \frac{ u_k k_\perp }{B_0 k_\|} \ll 1
\label{con1}
\eeq
(see \cite{Nazarenko2007}) where $|u_k|$ is the amplitude of the velocity field with wavenumbers $\sim {\bf k}$
(with $k_\|$ and $k_\perp$ the parallel and perpendicular projection of ${\bf k}$ to the magnetic field respectively).
$L_\|$ is the domain size in the direction of the field.
The inequality on the right implies sufficiently weak nonlinearity while the inequality on the left 
is needed for the presence of quasi resonances.
In this case WTT predicts that the energy spectrum is proportional to $k_\perp^{-2}$.
On the other hand if 
\beq
\frac{|u_k| k_\perp }{B_0 k_\| } \ll \frac{ 1 }{k_\perp L_\perp} \,\, \frac{1}{\sqrt{k_\| L_\|} } 
\label{con2} 
\eeq
the system becomes ``slaved" to the 2D modes $k_\|=0$ that evolve independently \cite{Nazarenko2007}.
Thus if the limit $B_0\to \infty$ is taken keeping the domain size $L$ fixed the system
becomes two dimensional \cite{Montgomery1981,Nazarenko2007}. Note that the two conditions (\ref{con1}),(\ref{con2}) 
allow the existence of an intermediate regime.

Here the large $B_0$ limit is explored further by means of numerical simulations.
Numerically, magneto-hydrodynamic (MHD) turbulence has been investigated by various groups
in the last decade \cite{Cho2000,Maron2001,Mason2008,Beresnyak2009,Beresnyak2010}.
The results of WTT were first demonstrated in \cite{Perez2007,Bigot2008} while the transition
to two dimensional dynamics has been investigated more recently in \cite{Bigot2011}.
In all these investigations the flow was forced in the largest scale of the system.
In this work the case where the system is forced in the small scales is explored 
and the possible development of an inverse cascade is examined.

An inverse energy cascade is known to exist in two-dimensional (2D) hydrodynamic turbulence \cite{Kraichnan67,Leith68,Batchelor69},
as a consequence of the conservation of vorticity. It results in a $k^{-5/3}$ spectrum for the large scales and a $k^{-3}$ spectrum
for the small scales. Strongly rotating flows 
\cite{Jacquin1990,Baroud2003,Mininni2009,Mininni2010,Thiele2009} 
or flows in thin boxes \cite{Celani2010} have been shown to have a dual cascade of energy with the large scale flow behaving like 
2D with an inverse cascade while the small scales being three dimensional with a direct cascade.
However, 2D MHD turbulence does not conserve the vorticity and energy
is cascading to small scales. On the contrary the square of the vector potential  is cascading to larger scales 
\cite{Fyfe1976,Pouquet1978}. 
For the system under investigation
we cannot a priori predict if the flow under the influence of a strong magnetic field 
will act as a 2D-{\it hydrodynamic} flow  by suppressing all magnetic fluctuations and thus have an inverse cascade;
or if magnetic fluctuations persist and the system will act as a 2D-MHD flow and thus 
not exhibit an  inverse cascade of energy. 
It is noted that a strong magnetic field is used very often to make flows of liquid metals behaving like two 
dimensional flows in experiments where an inverse cascade has been observed \cite{Sommeria1986,Tabeling1987}. These
flows however have very small magnetic Reynolds numbers and magnetic fluctuations are strongly suppressed.

%\cite{Iroshnikov1963}
%\cite{Kraichnan1965}
%\cite{Montgomery1981}
%\cite{Goldreich1995}
%\cite{Boldyrev2006}
%\cite{Alexakis2007}

%This paper is structured as follows. In the next section the problem under investigation is
%formulated and all measured quantities are defined. Section III presents the results from 9 runs,
%while in the last section the results are discussed in the more general framework of MHD turbulence.

%%%%%%%%%%%%%%%%%%%%%%%%%%%%%%%%%%%%%%%%%%%%%%%%%%%%%%%%%%%%%%%%%%%%%%%%%%%%%%%%%%%%%%%%%%%%%%%%%%%%%%%%%%%%%%%%%%%%%%%%%%%%%%%%
%%%%%%%%%%%%%%%%%%%%%%%%%%%%%%%%%%%%%%%%%%%%%%%%%%%%%%%%%%%%%%%%%%%%%%%%%%%%%%%%%%%%%%%%%%%%%%%%%%%%%%%%%%%%%%%%%%%%%%%%%%%%%%%
%%%%%%%%%%%%%%%%%%%%%%%%%%%%%%%%%%%%%%%%%%%%%%%%%%%%%%%%%%%%%%%%%%%%%%%%%%%%%%%%%%%%%%%%%%%%%%%%%%%%%%%%%%%%%%%%%%%%%%%%%%%%%%%%
%%%%%%%%%%%%%%%%%%%%%%%%%%%%%%%%%%%%%%%%%%%%%%%%%%%%%%%%%%%%%%%%%%%%%%%%%%%%%%%%%%%%%%%%%%%%%%%%%%%%%%%%%%%%%%%%%%%%%%%%%%%%%%%%
\section{formulation and numerical simulations}
%%%%%%%%%%%%%%%%%%%%%%%%%%%%%%%%%%%%%%%%%%%%%%%%%%%%%%%%%%%%%%%%%%%%%%%%%%%%%%%%%%%%%%%%%%%%%%%%%%%%%%%%%%%%%%%%%%%%%%%%%%%%%%%%
%%%%%%%%%%%%%%%%%%%%%%%%%%%%%%%%%%%%%%%%%%%%%%%%%%%%%%%%%%%%%%%%%%%%%%%%%%%%%%%%%%%%%%%%%%%%%%%%%%%%%%%%%%%%%%%%%%%%%%%%%%%%%%%%
%%%%%%%%%%%%%%%%%%%%%%%%%%%%%%%%%%%%%%%%%%%%%%%%%%%%%%%%%%%%%%%%%%%%%%%%%%%%%%%%%%%%%%%%%%%%%%%%%%%%%%%%%%%%%%%%%%%%%%%%%%%%%%%%
%%%%%%%%%%%%%%%%%%%%%%%%%%%%%%%%%%%%%%%%%%%%%%%%%%%%%%%%%%%%%%%%%%%%%%%%%%%%%%%%%%%%%%%%%%%%%%%%%%%%%%%%%%%%%%%%%%%%%%%%%%%%%%%%

We consider a flow of a conducting fluid inside a triple-periodic box of size $2\pi L$
in the presence of a strong guiding magnetic field $B_0$ in the ${\bf \hat{z}}$ direction.
The system is forced by a mechanical force ${\bf f}$ and an electro-motive force ${\bf \mathcal{E}}$.
The non-dimensional MHD equations then read:
%in the presence of a strong guiding magnetic field $B_0$ in the ${\bf \hat{z}}$ direction read:
\begin{eqnarray} 
\partial_t {\bf u} + {\bf u \cdot \nabla u} & =& V_{_A} \partial_z {\bf b} + {\bf b \cdot \nabla b} -\nabla P + G_{_K}^{^{-\frac{1}{2}}} \nabla^2 {\bf u} +               {\bf F}     \nonumber\\
%\label{MHD}\\
\partial_t {\bf b} + {\bf u \cdot \nabla b} & =& V_{_A} \partial_z {\bf u} + {\bf b \cdot \nabla u}     +G_{_M}^{^{-\frac{1}{2}}} \nabla^2 {\bf b} + M\nabla \times {\bf E},    \nonumber
\end{eqnarray}
where ${\bf u}$ is the velocity field and ${\bf b}$ is the magnetic field. 
Both fields are assumed to be solenoidal $\nabla \cdot {\bf u} = \nabla \cdot {\bf b} = 0$. 
${\bf F=f}/\|{\bf f}\|$ is the external mechanical force normalized to unit amplitude 
(where $\|\cdot\|$ stands for the $L_2$ norm).
${\bf E}$ is the external electro-motive force normalized so that its curl has unit amplitude ${\bf E}= {\bf \mathcal{E}} / \|\nabla \times {\bf \mathcal{E}} \| L$. 
%The domain that is considered in this work is a triple periodic box of size $2\pi L$.
%
The equations have been non-dimensionalized using the box size $L$ and the forcing amplitude $\| {\bf f}\|$.
With this choice four non-dimensional parameters appear.
$G_{_K}$ is the kinetic  Grasshof  number $G_{_K}\equiv \| {\bf f}\| L^3 /\nu^2  $, where $\nu$  is the viscosity.
$G_{_M}$ is the magnetic Grasshof  number $G_{_M}\equiv \| {\bf f}\| L^3 /\eta^2 $, where $\eta$ is the magnetic diffusivity.
In all the runs performed in this work $G_{_K}=G_{_M}$.
The amplitude of the external magnetic field relative to the forcing is controlled by the parameter 
$V_{_A}\equiv B_0 /\sqrt{\| {\bf f}\| L}$. Finally $M\equiv \| \nabla \times {\bf \mathcal{E}}\| / \| {\bf f}\|$ 
expresses the ratio of the electro-motive to the mechanical forcing.

A possible alternative to this non-dimensionalization choice would be to use the kinetic and magnetic Reynolds numbers,
typically defined as $Re \equiv \|{\bf u}\| L /\nu$ and $Rm\equiv \|{\bf u}\| L /\eta$ respectively. For our problem at hand however
where an inverse cascade is present it is not an attractive choice because the amplitude of the velocity is changing with time throughout 
the duration of the computation rendering $Re$ a function of time. 
%Furthermore a comparison of $Re$ between a flow with an inverse cascade  and flow with out will be misleading since a 

Both forcing mechanisms used in the numerical simulations consisted of a sum of Fourier modes with wave-numbers
inside a spherical shell ${\bf |k|}=k_f$.
The phases of the modes were changed randomly every time interval $\tau \sim \sqrt{L/\| {\bf f}\|}$. 
The forcing is isotropic in all runs except the last two that we discus in section \ref{sect}.
There was no averaged helicity or cross-helicity injected by the forcing by choosing:
$\langle {\bf F \cdot \nabla \times F} \rangle=0$ and $\langle {\bf F \cdot \nabla \times E} \rangle=0$. 
Since we are interested in the presence of an inverse cascade we are forcing relatively large wave numbers.

%%%%%%%%%%%%%%%%%%%%%%%%%%%%%%%%%%%%%%%%%%%%%%%%%%%%%%%%%%%%%%%%%%%%%%%%%%%%%%%%%%%%%%%%%%%%%%%%%%%%%%%%%%%%%%%%%%%%%%%%%%%%%%%%
%%%%%%%%%%%%%%%%%%%%%%%%%%%%%%%%%%%%%%%%%%%%%%%%%%%%%%%%%%%%%%%%%%%%%%%%%%%%%%%%%%%%%%%%%%%%%%%%%%%%%%%%%%%%%%%%%%%%%%%%%%%%%%%%
%%%%%%%%%%%%%%%%%%%%%%%%%%%%%%%%%%%%%%%%%%%%%%%%%%%%%%%%%%%%%%%%%%%%%%%%%%%%%%%%%%%%%%%%%%%%%%%%%%%%%%%%%%%%%%%%%%%%%%%%%%%%%%%%
%%%%%%%%%%%%%%%%%%%%%%%%%%%%%%%%%%%%%%%%%%%%%%%%%%%%%%%%%%%%%%%%%%%%%%%%%%%%%%%%%%%%%%%%%%%%%%%%%%%%%%%%%%%%%%%%%%%%%%%%%%%%%%%%
%\section{The Numerical simulations}
%%%%%%%%%%%%%%%%%%%%%%%%%%%%%%%%%%%%%%%%%%%%%%%%%%%%%%%%%%%%%%%%%%%%%%%%%%%%%%%%%%%%%%%%%%%%%%%%%%%%%%%%%%%%%%%%%%%%%%%%%%%%%%%%
%%%%%%%%%%%%%%%%%%%%%%%%%%%%%%%%%%%%%%%%%%%%%%%%%%%%%%%%%%%%%%%%%%%%%%%%%%%%%%%%%%%%%%%%%%%%%%%%%%%%%%%%%%%%%%%%%%%%%%%%%%%%%%%%
%%%%%%%%%%%%%%%%%%%%%%%%%%%%%%%%%%%%%%%%%%%%%%%%%%%%%%%%%%%%%%%%%%%%%%%%%%%%%%%%%%%%%%%%%%%%%%%%%%%%%%%%%%%%%%%%%%%%%%%%%%%%%%%%
%%%%%%%%%%%%%%%%%%%%%%%%%%%%%%%%%%%%%%%%%%%%%%%%%%%%%%%%%%%%%%%%%%%%%%%%%%%%%%%%%%%%%%%%%%%%%%%%%%%%%%%%%%%%%%%%%%%%%%%%%%%%%%%%

The MHD equations  were solved using a standard pseudo-spectral method and a third order in time Runge-Kuta 
\cite{Minini_code1,Minini_code2}. The resolution used in all the runs was $512^3$ grid points. 
%$G_{_K}$ was chosen so that each run has a well resolved spectrum. 
In the 9 different runs that were performed the amplitude of the external magnetic field, and the way the system is forced 
was varied. The table \ref{tbl}  gives all the parameters of the runs.
%%%%%%%%%%%%%%%%%%%%%%%%%%%%%%%%%%%%%%%%%%%%%%%%%%%%%%%%%%%%%%%%%%%%%%%%%%%%%%%%%%%%%%%%%%%%%%%%%%%%%%%%%%%%%%%%%%%
%%%%%%%%%%%%%%%%%%%%%%%%%%%%%%%%%%%%%%%%%%%%%%%%%%%%%%%%%%%%%%%%%%%%%%%%%%%%%%%%%%%%%%%%%%%%%%%%%%%%%%%%%%%%%%%%%%%
\begin{table}
\label{tbl}
\begin{tabular}{| c      |     c      |      c        |   c         |   c         |    c                    | }
\hline
                 RUNS    &$\quad G_{_K}^{^{\frac{1}{2}}}/10^3 \quad $ & $\quad V_{_A} \quad $   &    $\quad k_fL\quad$    &  $\quad M_0 \quad $    &  Isotropy        \\
\hline \hline
    R1   &  5.0           &   {    5.0 }  &  {    8-10} &  {    0.0}  &    I                   \\
    R2   &  2.5           &   {\bf 2.0}   &       8-10  &       0.0   &    I                   \\
    R3   &  2.5           &   {\bf 10.0}  &       8-10  &       0.0   &    I                   \\
    R4   & 10.0           &      5.0      & {\bf  4-5 } &       0.0   &    I                   \\
    R5   &  2.5           &      5.0      & {\bf 16-20} &       0.0   &    I                   \\
    R6   &  3.3           &      5.0      &       8-10  &  {\bf 0.4}  &    I                   \\
    R7   &  2.5           &      5.0      &       8-10  &  {\bf 0.6}  &    I                   \\
    R8   &  2.5           &      5.0      &       8-10  &       0.0   & $\,\,{\bf F(k_\|,0)=0   }\,\,$ \\
    R9   & 16.0           &      5.0      &       8-10  &       0.0   & $\,\,{\bf F(0,k_\perp)=0}\,\,$ \\
\hline
\end{tabular}
\caption{Table with the parameters of all runs. ``I" in the last column stands for isotropic forcing.
With boldface are marked the parameters that are varied with respect to R1. \label{table1}}
\end{table}

The choice of $G_{_K}$ in most runs is rather conservative because it is not known beforehand what effect on the resolution
requirements a change in each of the parameters has. In any case, in all runs a well resolved spectrum was observed.
The last run (R9) has a large value of $G_{_K}$ because in this case
for similar forcing amplitude with the other runs the flow is less efficient in absorbing energy because the $k_\|=0$ modes 
are not forced. In this run also the time scale of the forcing was set to $\tau \sim (B_0 k_f)^{-1}$ to be closer in resonance
with the forced Alfven-modes and to improve this absorption efficiency. 

The diagnostics used are based on energy spectra and energy fluxes that are now defined.
If ${\bf \hat{u}_k}$ and ${\bf \hat{b}_k}$ are the Fourier modes of the velocity and magnetic field of wavenumber ${\bf k}$
then the two dimensional energy spectra $E_u(k_\perp,k_{\|})$ and $E_b(k_\perp,k_{\|})$  are defined as
\[ E_u(k_\perp,k_{\|})= \frac{1}{2} \sum {\bf |\hat{u}_k}|^2 , \quad E_b(k_\perp,k_{\|})= \frac{1}{2} \sum {\bf |\hat{b}_k}|^2\]
where the sum is restricted in the wavenumbers
$k_{\|} \le |k_z| < k_{\|}+1$ and $k_\perp \le \sqrt{k_x^2+k_y^2} < k_\perp+1$.
%Similarly the magnetic energy spectrum $E_b(k_\perp,k_{\|})$ is defined.
%
The averaged energy spectra in the parallel direction are then defined as
\[ \bar{E}_u(k_\perp)=\sum_{k_{\|}} E_u(k_\perp,k_{\|}), \quad \bar{E}_b(k_\perp)=\sum_{k_{\|}} E_b(k_\perp,k_{\|}).\]
%and  similarly $\bar{E}_b(k_\perp)$.
%
The total kinetic energy $E_{_K}$ and magnetic energy $E_{_M}$ are then given by
\[
E_{_K}= \sum_{k_\perp} \bar{E}_u(k_\perp),\quad E_{_M}= \sum_{k_\perp} \bar{E}_b(k_\perp).
\]

Since the development of a direct or an inverse cascade is  expected to be anisotropic
% of interest in this work 
we need to define the flux of energy through an arbitrary surface in Fourier space.
If $\bf u_{_D}$ and $\bf b_{_D}$ stand for the projection of the two fields ${\bf u,b}$ to the flows whose Fourier 
transform contain modes only inside the Fourier domain ``D", the energy flux through this domain is given by
%\beq 
\[
\Pi_{_D} = \int [{\bf u_{_D} u\cdot \nabla u } - {\bf u_{_D} b\cdot \nabla b } 
               + {\bf b_{_D} u\cdot \nabla b } - {\bf b_{_D} b\cdot \nabla u }   ]      dV.
\]
%\eeq
See \cite{Alexakis2007} for more details.
In this work we will consider the flux through cylinders and planes.
By $\Pi_\perp(k_\perp)$ we will refer to the flux of energy through a cylinder of radius $\sqrt{k_x^2+k_y^2}=k_\perp$
and by $\Pi_\perp(k_{\|})$  we will refer to the flux of energy through the planes $|k_z|=k_{\|}$. 
Positive flux implies cascade of energy to the small scales while a negative flux implies cascade to the large scales.

%%%%%%%%%%%%%%%%%%%%%%%%%%%%%%%%%%%%%%%%%%%%%%%%%%%%%%%%%%%%%%%%%%%%%%%%%%%%%%%%%%%%%%%%%%%%%%%%%%%%%%%%%%%%%%%%%%%%%%%%%%%%%%%
%%%%%%%%%%%%%%%%%%%%%%%%%%%%%%%%%%%%%%%%%%%%%%%%%%%%%%%%%%%%%%%%%%%%%%%%%%%%%%%%%%%%%%%%%%%%%%%%%%%%%%%%%%%%%%%%%%%%%%%%%%%%%%%
%%%%%%%%%%%%%%%%%%%%%%%%%%%%%%%%%%%%%%%%%%%%%%%%%%%%%%%%%%%%%%%%%%%%%%%%%%%%%%%%%%%%%%%%%%%%%%%%%%%%%%%%%%%%%%%%%%%%%%%%%%%%%%%
%%%%%%%%%%%%%%%%%%%%%%%%%%%%%%%%%%%%%%%%%%%%%%%%%%%%%%%%%%%%%%%%%%%%%%%%%%%%%%%%%%%%%%%%%%%%%%%%%%%%%%%%%%%%%%%%%%%%%%%%%%%%%%%
%%%%%%%%%%%%%%%%%%%%%%%%%%%%%%%%%%%%%%%%%%%%%%%%%%%%%%%%%%%%%%%%%%%%%%%%%%%%%%%%%%%%%%%%%%%%%%%%%%%%%%%%%%%%%%%%%%%%%%%%%%%%%%%
%%%%%%%%%%%%%%%%%%%%%%%%%%%%%%%%%%%%%%%%%%%%%%%%%%%%%%%%%%%%%%%%%%%%%%%%%%%%%%%%%%%%%%%%%%%%%%%%%%%%%%%%%%%%%%%%%%%%%%%%%%%%%%%
\section{Results}
%%%%%%%%%%%%%%%%%%%%%%%%%%%%%%%%%%%%%%%%%%%%%%%%%%%%%%%%%%%%%%%%%%%%%%%%%%%%%%%%%%%%%%%%%%%%%%%%%%%%%%%%%%%%%%%%%%%%%%%%%%%%%%%
%%%%%%%%%%%%%%%%%%%%%%%%%%%%%%%%%%%%%%%%%%%%%%%%%%%%%%%%%%%%%%%%%%%%%%%%%%%%%%%%%%%%%%%%%%%%%%%%%%%%%%%%%%%%%%%%%%%%%%%%%%%%%%%
%%%%%%%%%%%%%%%%%%%%%%%%%%%%%%%%%%%%%%%%%%%%%%%%%%%%%%%%%%%%%%%%%%%%%%%%%%%%%%%%%%%%%%%%%%%%%%%%%%%%%%%%%%%%%%%%%%%%%%%%%%%%%%%
%%%%%%%%%%%%%%%%%%%%%%%%%%%%%%%%%%%%%%%%%%%%%%%%%%%%%%%%%%%%%%%%%%%%%%%%%%%%%%%%%%%%%%%%%%%%%%%%%%%%%%%%%%%%%%%%%%%%%%%%%%%%%%%
%%%%%%%%%%%%%%%%%%%%%%%%%%%%%%%%%%%%%%%%%%%%%%%%%%%%%%%%%%%%%%%%%%%%%%%%%%%%%%%%%%%%%%%%%%%%%%%%%%%%%%%%%%%%%%%%%%%%%%%%%%%%%%%
%%%%%%%%%%%%%%%%%%%%%%%%%%%%%%%%%%%%%%%%%%%%%%%%%%%%%%%%%%%%%%%%%%%%%%%%%%%%%%%%%%%%%%%%%%%%%%%%%%%%%%%%%%%%%%%%%%%%%%%%%%%%%%%

%%%%%%%%%%%%%%%%%%%%%%%%%%%%%%%%%%%%%%%%%%%%%%%%%%%%%%%%%%%%%%%%%%%%%%%%%%%%%%%%%%%%%%%%%%%%%%%%%%%%%%%%%%%%%%%%%%%%%%%%%%%%%%%
%%%%%%%%%%%%%%%%%%%%%%%%%%%%%%%%%%%%%%%%%%%%%%%%%%%%%%%%%%%%%%%%%%%%%%%%%%%%%%%%%%%%%%%%%%%%%%%%%%%%%%%%%%%%%%%%%%%%%%%%%%%%%%%
\subsection{A pilot run }%%%%%%%%%%%%%%%%%%%%%%%%%%%%%%%%%%%%%%%%%%%%%%%%%%%%%%%%%%%%%%%%%%%%%%%%%%%%%%%%%%%%%%%%%%%%%%%%%%%%%%
%%%%%%%%%%%%%%%%%%%%%%%%%%%%%%%%%%%%%%%%%%%%%%%%%%%%%%%%%%%%%%%%%%%%%%%%%%%%%%%%%%%%%%%%%%%%%%%%%%%%%%%%%%%%%%%%%%%%%%%%%%%%%%%
%%%%%%%%%%%%%%%%%%%%%%%%%%%%%%%%%%%%%%%%%%%%%%%%%%%%%%%%%%%%%%%%%%%%%%%%%%%%%%%%%%%%%%%%%%%%%%%%%%%%%%%%%%%%%%%%%%%%%%%%%%%%%%%

%%%%%%%%%%%%%%%%%%%%%%%%%%%%%%%%%%%%%%%%%%%%%%%%%%%%%%%%%%%%%%%%%%%%%%%%%%%%%%%%%%%%%%%%%%%%%%%%%%%%%%%%%%%%%%%%%%%%%%%%%%%%%%%
\begin{figure}                                                                                                                %
\includegraphics[width=8cm]{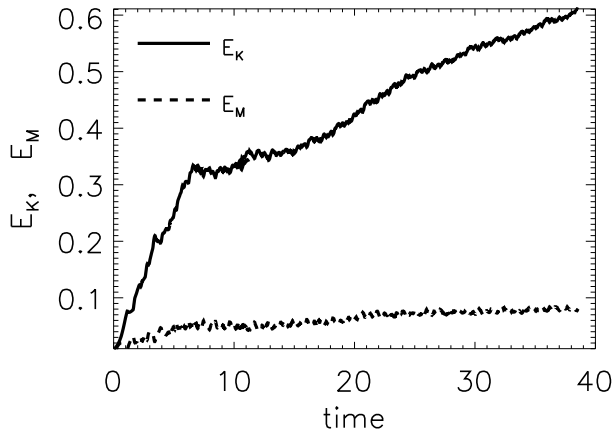}                                                                                       %
\caption{\label{A1} Time evolution of the kinetic energy $E_{_K}$ (solid line) and the magnetic energy $E_{_M}$ (dashed line) %
for run R1.}                                                                                                                  %
\end{figure}                                                                                                                  %  
%%%%%%%%%%%%%%%%%%%%%%%%%%%%%%%%%%%%%%%%%%%%%%%%%%%%%%%%%%%%%%%%%%%%%%%%%%%%%%%%%%%%%%%%%%%%%%%%%%%%%%%%%%%%%%%%%%%%%%%%%%%%%%% 

%%%%%%%%%%%%%%%%%%%%%%%%%%%%%%%%%%%%%%%%%%%%%%%%%%%%%%%%%%%%%%%%%%%%%%%%%%%%%%%%%%%%%%%%%%%%%%%%%%%%%%%%%%%%%%%%%%%%%%%%%%%%%%%
\begin{figure}                                                                                                                %
\includegraphics[width=8cm]{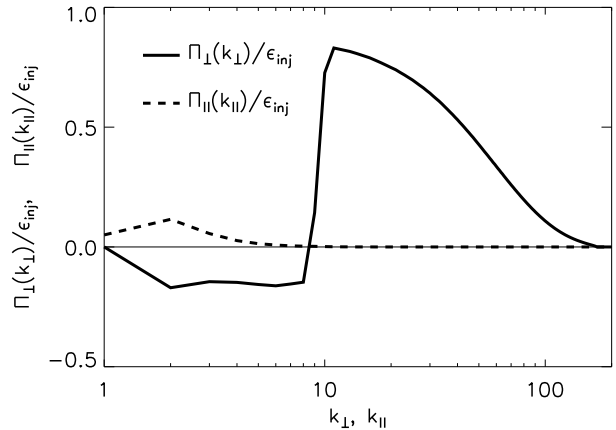}                                                                                       %
\caption{\label{A2}The energy flux for run R1 in the perpendicular direction (solid line)                                     %
          and parallel direction (dashed line).}                                                                              % 
\end{figure}                                                                                                                  %  
%%%%%%%%%%%%%%%%%%%%%%%%%%%%%%%%%%%%%%%%%%%%%%%%%%%%%%%%%%%%%%%%%%%%%%%%%%%%%%%%%%%%%%%%%%%%%%%%%%%%%%%%%%%%%%%%%%%%%%%%%%%%%%% 

%%%%%%%%%%%%%%%%%%%%%%%%%%%%%%%%%%%%%%%%%%%%%%%%%%%%%%%%%%%%%%%%%%%%%%%%%%%%%%%%%%%%%%%%%%%%%%%%%%%%%%%%%%%%%%%%%%%%%%%%%%%%%%%
\begin{figure}                                                                                                                %
\includegraphics[width=8cm]{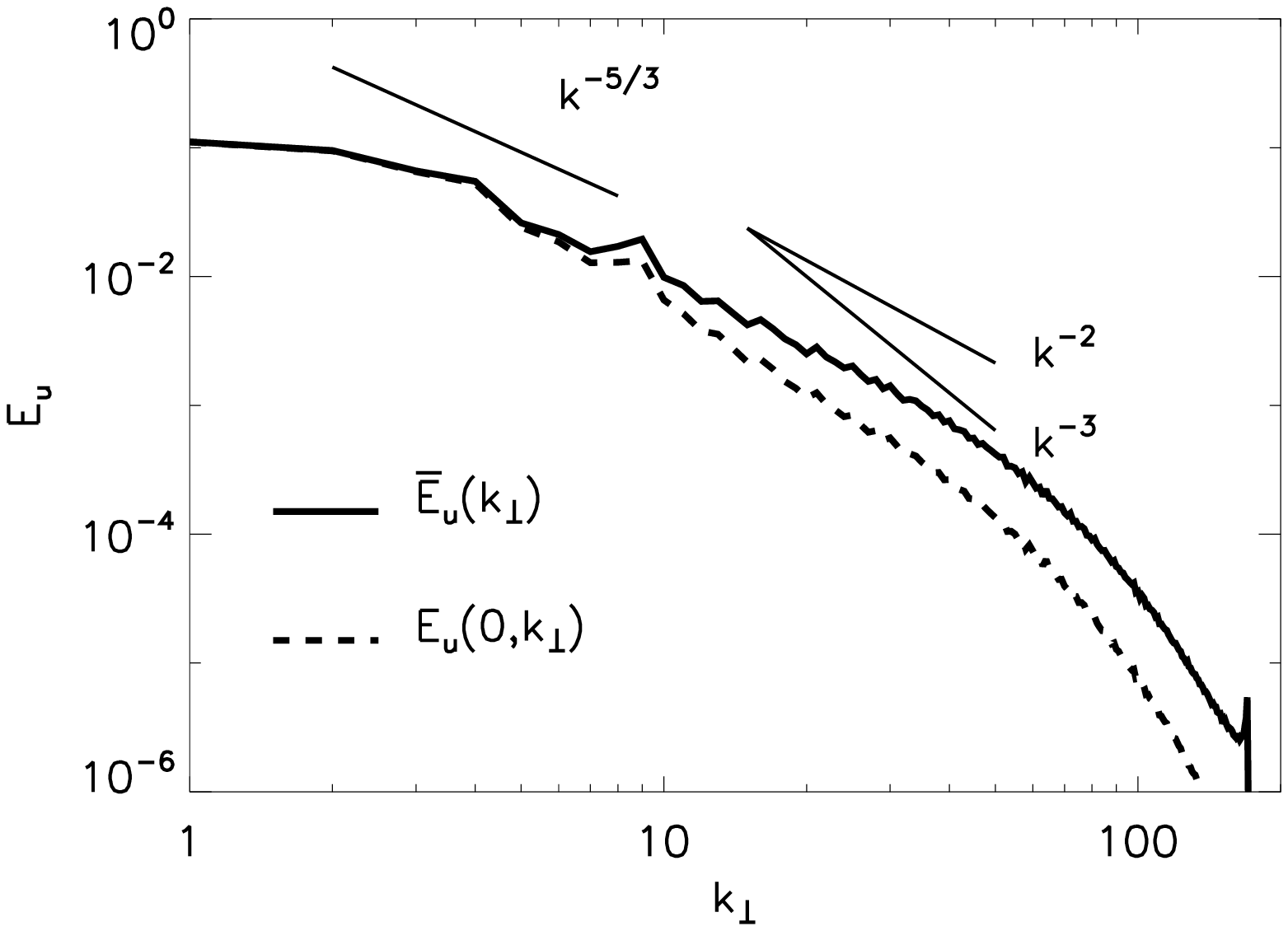}      \\                                                                              %
\includegraphics[width=8cm]{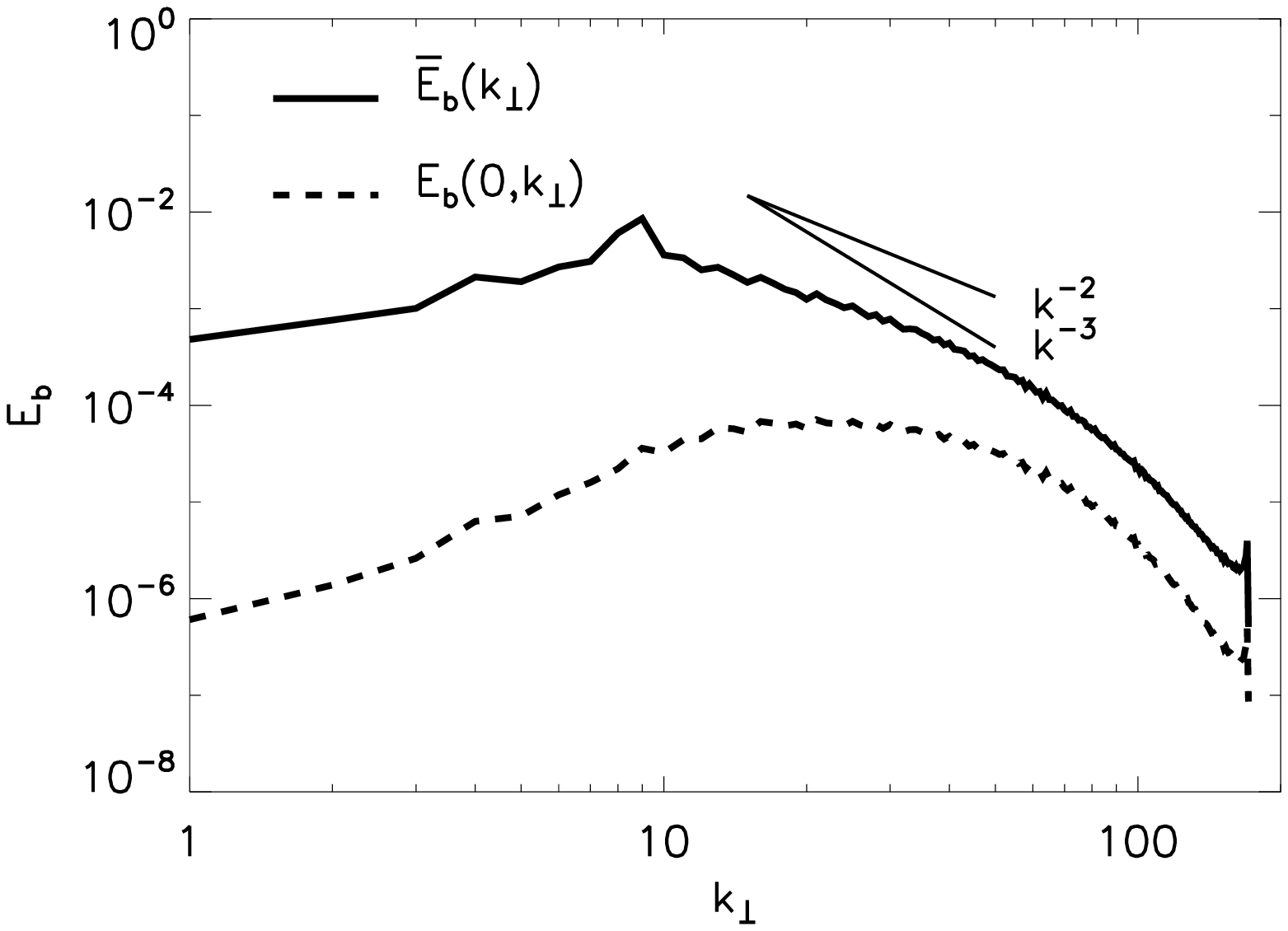}                                                                                      %
\caption{\label{A3} Kinetic (top panel) and magnetic (bottom panel)  energy spectra of run R1.                                %
The solid lines correspond to the averaged spectra $\bar{E}_u(k_\perp),\bar{E}_b(k_\perp) $ while the dashed lines            %
indicate the zeroth component ($k_\|=0$) of the two dimensional energy spectra $E_u(0,k_\perp)$ and $E_b(0,k_\perp)$.         %
The straight lines show for reference the power law spectra $k^{-5/3},k^{-2},k^{-3}$.}                                        %
\end{figure}                                                                                                                  %
%%%%%%%%%%%%%%%%%%%%%%%%%%%%%%%%%%%%%%%%%%%%%%%%%%%%%%%%%%%%%%%%%%%%%%%%%%%%%%%%%%%%%%%%%%%%%%%%%%%%%%%%%%%%%%%%%%%%%%%%%%%%%%%

The first run in table \ref{table1} serves as a basic run to which all other runs are compared.
For this reason this run is examined in more detail. Figure \ref{A1} shows the evolution of the kinetic and magnetic energy 
as a function of time. As can be seen the magnetic energy grows and saturates very fast at a relatively small amplitude. The
kinetic energy on the other hand, after an initial fast growth  transitions to a slower 
increasing phase. Up until the end of the numerical simulation this slow growth persists. 
The reason for this growth is the inverse cascade of the kinetic energy that accumulates energy in the large scales. 

This inverse cascade is demonstrated more clearly in figure \ref{A2}. This figure shows
the parallel and perpendicular energy flux normalized by the total energy injection rate. 
The perpendicular energy flux is positive for wavenumbers larger than the forcing wavenumber (direct cascade) 
while for smaller wave numbers a negative constant energy flux (inverse cascade) can be seen. 
On the other hand the parallel energy flux, shown by the dashed line, is small and always positive ({\it ie} direct).
 
The presence of an inverse cascade can also indicated by looking at the energy spectra at late times. 
The top panel of figure \ref{A3} shows the kinetic energy spectrum $\bar{E}_u$ of run R1 averaged over 
several outputs close to the end of the simulation. It can be clearly seen that most of the energy is 
concentrated in the large scales. The dashed line in this panel shows $E_u(0,k_\perp)$. 
At large scales this line is identical to the  $\bar{E}_u$ spectrum thus the energy in these scales is mostly contained
in the 2D-modes $k_\|=0$. This means that the flow in the large scales is almost 2D.
(Here the flow is referred to as 2D in the sense that $u$ has no dependence on the $z$-direction and 
not that the $u_z$ component is absent.)
On the other hand at the small scales  $E_u(0,k_\perp)$ is significantly smaller than
$\bar{E}_u$ thus the 2D-modes contain only a small fraction of the energy and therefor the flow is three-dimensional.

The bottom panel of fig \ref{A3} compares the magnetic energy spectra $E_b(0,k_\perp)$ and $\bar{E}_b $. 
Unlike the velocity field the magnetic field remains strongly three-dimensional 
for all scales since $E_b(0,k_\perp) \ll \bar{E}_b $. 
The amplitude of the magnetic energy is much smaller than that of 
the kinetic energy in the large scales but of the same order in the small scales.  
This is essential for the presence of the 2D-inverse cascade. If the magnetic field fluctuations were strong enough in the
large scales the flow would behave as a 2D-MHD flow with a direct cascade.

The $k^{-5/3}$ scaling prediction for the 2D inverse cascade, the $k^{-3}$ for the direct 2D cascade and the 
$k^{-2}$ prediction of WTT are shown as a reference.
The observed spectra are compatible with $k^{-5/3}$ in the large scales and $k^{-2}$ in the small scales however
the inertial ranges in the examined flow are too small to be conclusive.
 
%%%%%%%%%%%%%%%%%%%%%%%%%%%%%%%%%%%%%%%%%%%%%%%%%%%%%%%%%%%%%%%%%%%%%%%%%%%%%%%%%%%%%%%%%%%%%%%%%%%%%%%%%%%%%%%%%%%%%%%%%%%%%%%
%%%%%%%%%%%%%%%%%%%%%%%%%%%%%%%%%%%%%%%%%%%%%%%%%%%%%%%%%%%%%%%%%%%%%%%%%%%%%%%%%%%%%%%%%%%%%%%%%%%%%%%%%%%%%%%%%%%%%%%%%%%%%%%
\subsection{Guiding magnetic field strength }%%%%%%%%%%%%%%%%%%%%%%%%%%%%%%%%%%%%%%%%%%%%%%%%%%%%%%%%%%%%%%%%%%%%%%%%%%%%%%%%%%
%%%%%%%%%%%%%%%%%%%%%%%%%%%%%%%%%%%%%%%%%%%%%%%%%%%%%%%%%%%%%%%%%%%%%%%%%%%%%%%%%%%%%%%%%%%%%%%%%%%%%%%%%%%%%%%%%%%%%%%%%%%%%%%
%%%%%%%%%%%%%%%%%%%%%%%%%%%%%%%%%%%%%%%%%%%%%%%%%%%%%%%%%%%%%%%%%%%%%%%%%%%%%%%%%%%%%%%%%%%%%%%%%%%%%%%%%%%%%%%%%%%%%%%%%%%%%%%

%%%%%%%%%%%%%%%%%%%%%%%%%%%%%%%%%%%%%%%%%%%%%%%%%%%%%%%%%%%%%%%%%%%%%%%%%%%%%%%%%%%%%%%%%%%%%%%%%%%%%%%%%%%%%%%%%%%%%%%%%%%%%%%
\begin{figure}                                                                                                                %
\includegraphics[width=8cm]{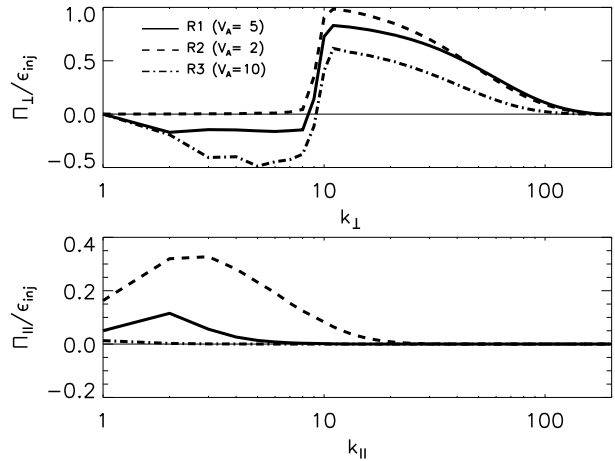}                                                                                       %
\caption{\label{B1} Top panel: The energy flux in the perpendicular direction for R1 ($V_{_A}=5$,  solid line),               %
                                                                                  R2 ($V_{_A}=2$,  dashed line),              %
                                                                                  R3 ($V_{_A}=10$, dashed-dot line).          %
                    Bottom panel: The energy flux in the perpendicular direction for the same runs.}                          %
\end{figure}                                                                                                                  %
%%%%%%%%%%%%%%%%%%%%%%%%%%%%%%%%%%%%%%%%%%%%%%%%%%%%%%%%%%%%%%%%%%%%%%%%%%%%%%%%%%%%%%%%%%%%%%%%%%%%%%%%%%%%%%%%%%%%%%%%%%%%%%%

%%%%%%%%%%%%%%%%%%%%%%%%%%%%%%%%%%%%%%%%%%%%%%%%%%%%%%%%%%%%%%%%%%%%%%%%%%%%%%%%%%%%%%%%%%%%%%%%%%%%%%%%%%%%%%%%%%%%%%%%%%%%%%%
\begin{figure}                                                                                                                %
\includegraphics[width=8cm]{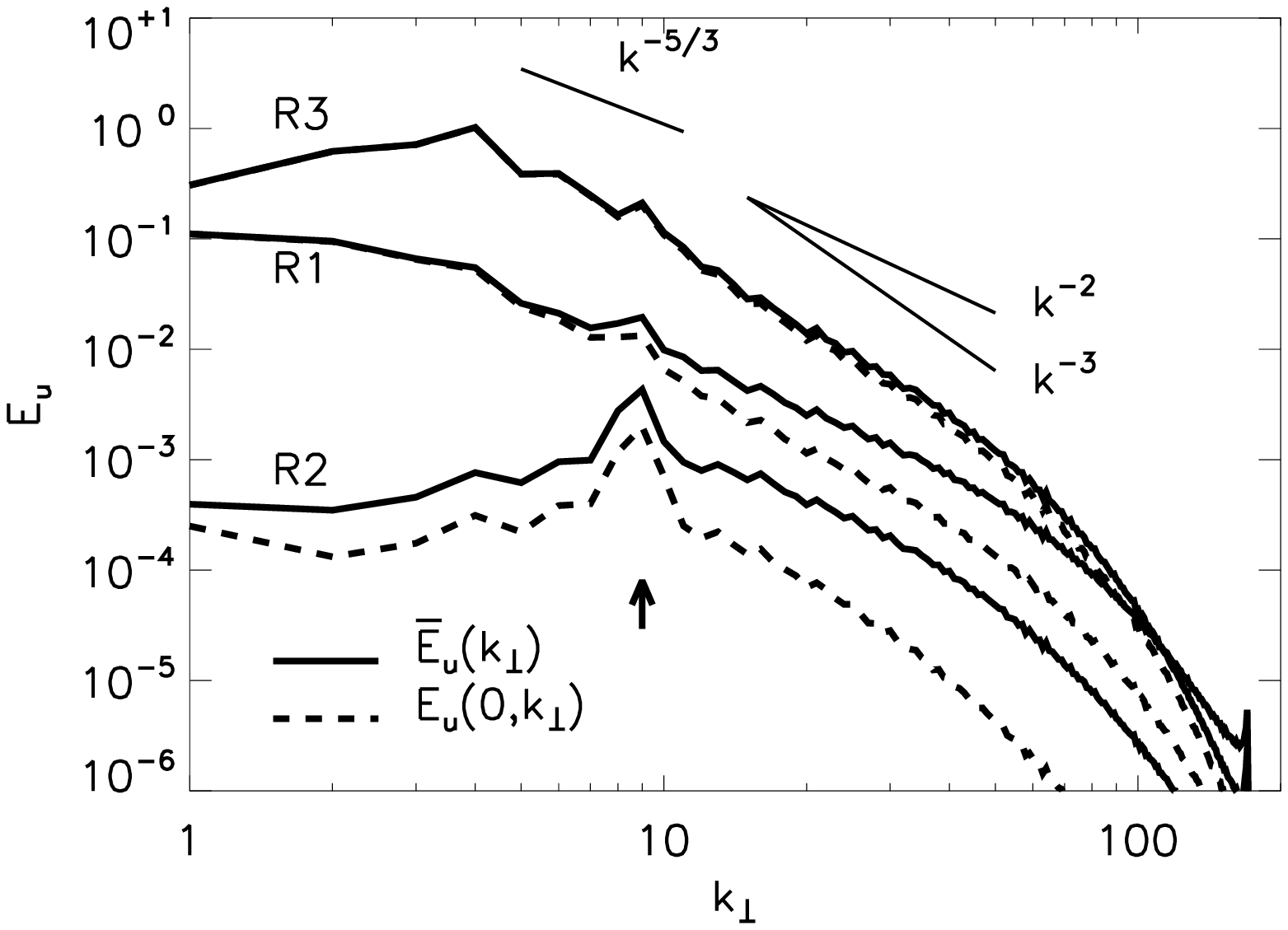}                                 \\                                         %
\includegraphics[width=8cm]{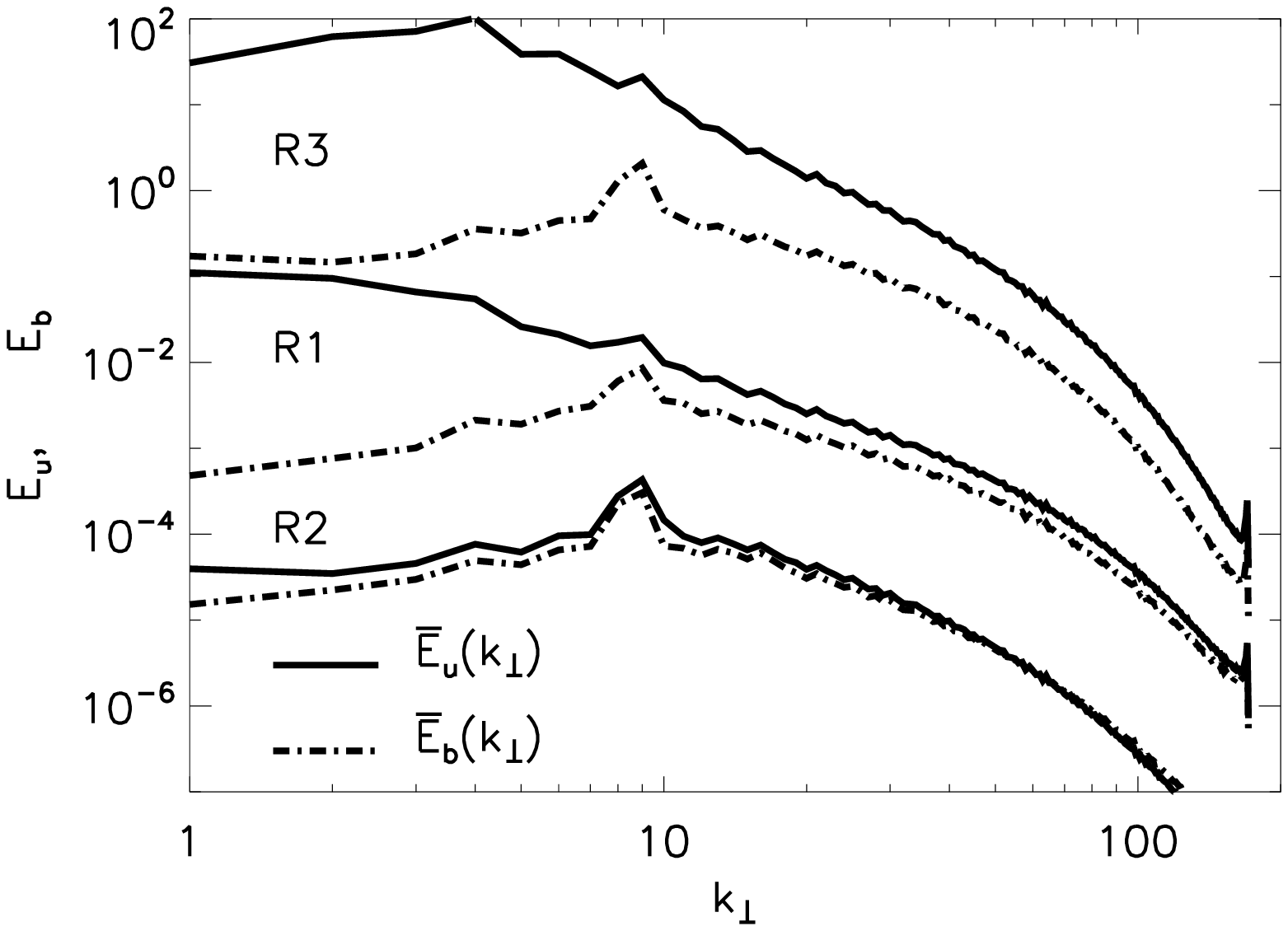}                                                                            %
\caption{\label{B2} Top panel: The  kinetic energy spectra of $\bar{E}_u(k_\perp)$ (solid line),                              %
                                                          and      $E_u(0,k_\perp)$ (dashed line)                             %
                 for R3, $V_{_A}=10$ (top lines), R1, $V_{_A}=5$ (middle lines), R2, $V_{_A}=2$ (bottom lines).               %
                 Bottom panel: The kinetic energy spectra $\bar{E}_u(k_\perp)$ (solid line),                                  %
                 compared to the magnetic energy spectra $\bar{E}_b(k_\perp)$ (dashed line)                                   %
                 of the same runs and with the same order.                                                                    %
The spectra have been shifted for reasons of clarity.    }                                                                    %
\end{figure}                                                                                                                  %
%%%%%%%%%%%%%%%%%%%%%%%%%%%%%%%%%%%%%%%%%%%%%%%%%%%%%%%%%%%%%%%%%%%%%%%%%%%%%%%%%%%%%%%%%%%%%%%%%%%%%%%%%%%%%%%%%%%%%%%%%%%%%%%

As a next step the dependence of the inverse cascade, observed in R1, on the amplitude of the uniform 
magnetic field is examined. Runs R2 and R3 have all parameters similar to run R1 but different value of 
the magnetic field amplitude.
The flux of energy in both directions for the runs R1,R2 and R3 are compared in figure \ref{B1}. 
As expected the amplitude of the uniform magnetic field has a drastic effect on the energy flux. 
The top panel of this figure shows $\Pi_\perp(k_\perp)$. R2 (dashed line) that has smaller value of $V_{_A}$ than run R1
(solid line) has no inverse cascade and a stronger direct cascade. 
R3 (dashed-dot line) that has larger value of $V_{_A}$ has on the contrary a stronger
inverse cascade and a weaker forward cascade. 
The bottom panel of figure \ref{B1} shows the energy flux in the parallel direction. As the magnetic field is increased the 
flux to large $k_z$ is decreased. This expected since in the $V_{_A}=\infty$ limit there is cascade only in perpendicular
direction.

The spectra for these runs are compared in figure \ref{B2}. The top panel of this figure shows  the kinetic energy spectra 
$\bar{E}_u(k_\perp)$ and $E_u(0,k_\perp)$. The spectra have been shifted for reasons of clarity. 
%(by the same factor for each run). 
In the two runs R1 and R3  that showed an inverse cascade, energy is concentrated 
in the largest scales. What can also be observed is that as $V_{_A}$ is increased the flow comes closer to a
two dimensional flow. For R2 for which $V_{_A}=2$ and no inverse cascade is observed the flow is far from two dimensional
even at the largest scales. 

The bottom panel panel of figure \ref{B2} compares the kinetic energy spectra $\bar{E}_u(k_\perp)$ (solid line) with
the magnetic energy spectra $\bar{E}_b(k_\perp)$ (dashed line). As the uniform magnetic field is increased
the magnetic fluctuations are decreased compared to the velocity fluctuations. Note that for R3 the magnetic fluctuations are
almost negligible in all scales while for R2 the two fluctuating fields are in equipartition. 
A possible interpretation of for this behavior is the following.  Since in these runs there is no forcing for the magnetic field 
the magnetic fluctuations can be generated only by the stretching of field lines of the uniform component or by a dynamo
mechanism. However since the flow comes close to a 2D flow as $V_{_A}$ is increased neither of these mechanisms is possible. 
The dynamo mechanism however could depend on the magnetic Reynolds number that for these runs it is relatively small.

%%%%%%%%%%%%%%%%%%%%%%%%%%%%%%%%%%%%%%%%%%%%%%%%%%%%%%%%%%%%%%%%%%%%%%%%%%%%%%%%%%%%%%%%%%%%%%%%%%%%%%%%%%%%%%%%%%%%%%%%%%%%%%%
%%%%%%%%%%%%%%%%%%%%%%%%%%%%%%%%%%%%%%%%%%%%%%%%%%%%%%%%%%%%%%%%%%%%%%%%%%%%%%%%%%%%%%%%%%%%%%%%%%%%%%%%%%%%%%%%%%%%%%%%%%%%%%%
\subsection{Forcing scale}%%%%%%%%%%%%%%%%%%%%%%%%%%%%%%%%%%%%%%%%%%%%%%%%%%%%%%%%%%%%%%%%%%%%%%%%%%%%%%%%%%%%%%%%%%%%%%%%%%%%%
%%%%%%%%%%%%%%%%%%%%%%%%%%%%%%%%%%%%%%%%%%%%%%%%%%%%%%%%%%%%%%%%%%%%%%%%%%%%%%%%%%%%%%%%%%%%%%%%%%%%%%%%%%%%%%%%%%%%%%%%%%%%%%%
%%%%%%%%%%%%%%%%%%%%%%%%%%%%%%%%%%%%%%%%%%%%%%%%%%%%%%%%%%%%%%%%%%%%%%%%%%%%%%%%%%%%%%%%%%%%%%%%%%%%%%%%%%%%%%%%%%%%%%%%%%%%%%%

%%%%%%%%%%%%%%%%%%%%%%%%%%%%%%%%%%%%%%%%%%%%%%%%%%%%%%%%%%%%%%%%%%%%%%%%%%%%%%%%%%%%%%%%%%%%%%%%%%%%%%%%%%%%%%%%%%%%%%%%%%%%%%%
\begin{figure}                                                                                                                %
\includegraphics[width=8cm]{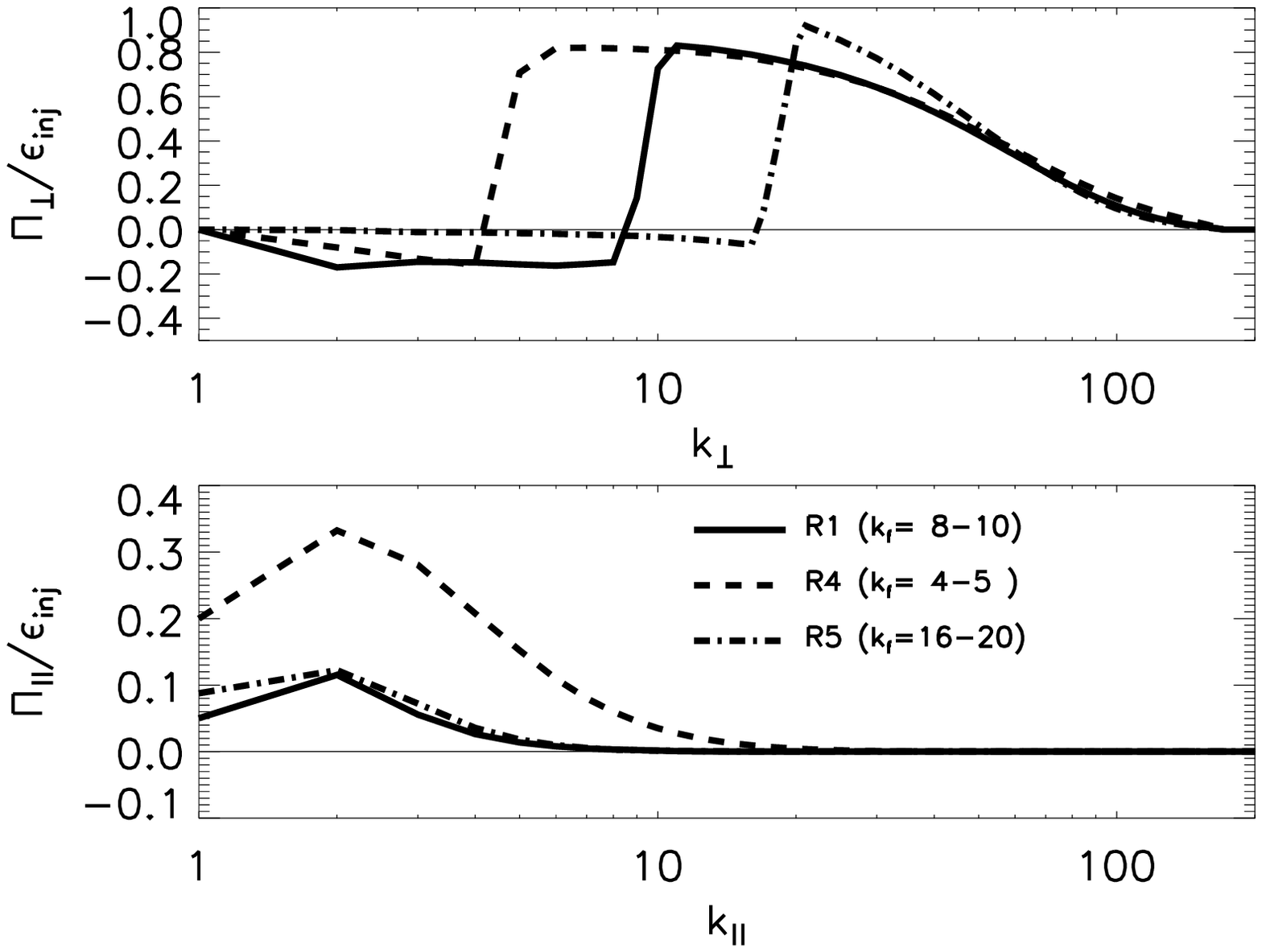}                                                                                       %
\caption{\label{C1} Top panel: The energy flux in the perpendicular direction for R1 ($8<k_f\le10$,  solid line),             %
                                                                                  R4 ($4<k_k\le5$,   dashed line),            %
                                                                                  R5 ($16<k_f\le20$, dashed-dot line).        %
                    Bottom panel: The energy flux in the perpendicular direction for the same runs.}                          %
\end{figure}                                                                                                                  %
%%%%%%%%%%%%%%%%%%%%%%%%%%%%%%%%%%%%%%%%%%%%%%%%%%%%%%%%%%%%%%%%%%%%%%%%%%%%%%%%%%%%%%%%%%%%%%%%%%%%%%%%%%%%%%%%%%%%%%%%%%%%%%%

%%%%%%%%%%%%%%%%%%%%%%%%%%%%%%%%%%%%%%%%%%%%%%%%%%%%%%%%%%%%%%%%%%%%%%%%%%%%%%%%%%%%%%%%%%%%%%%%%%%%%%%%%%%%%%%%%%%%%%%%%%%%%%%
\begin{figure}                                                                                                                %
\includegraphics[width=8cm]{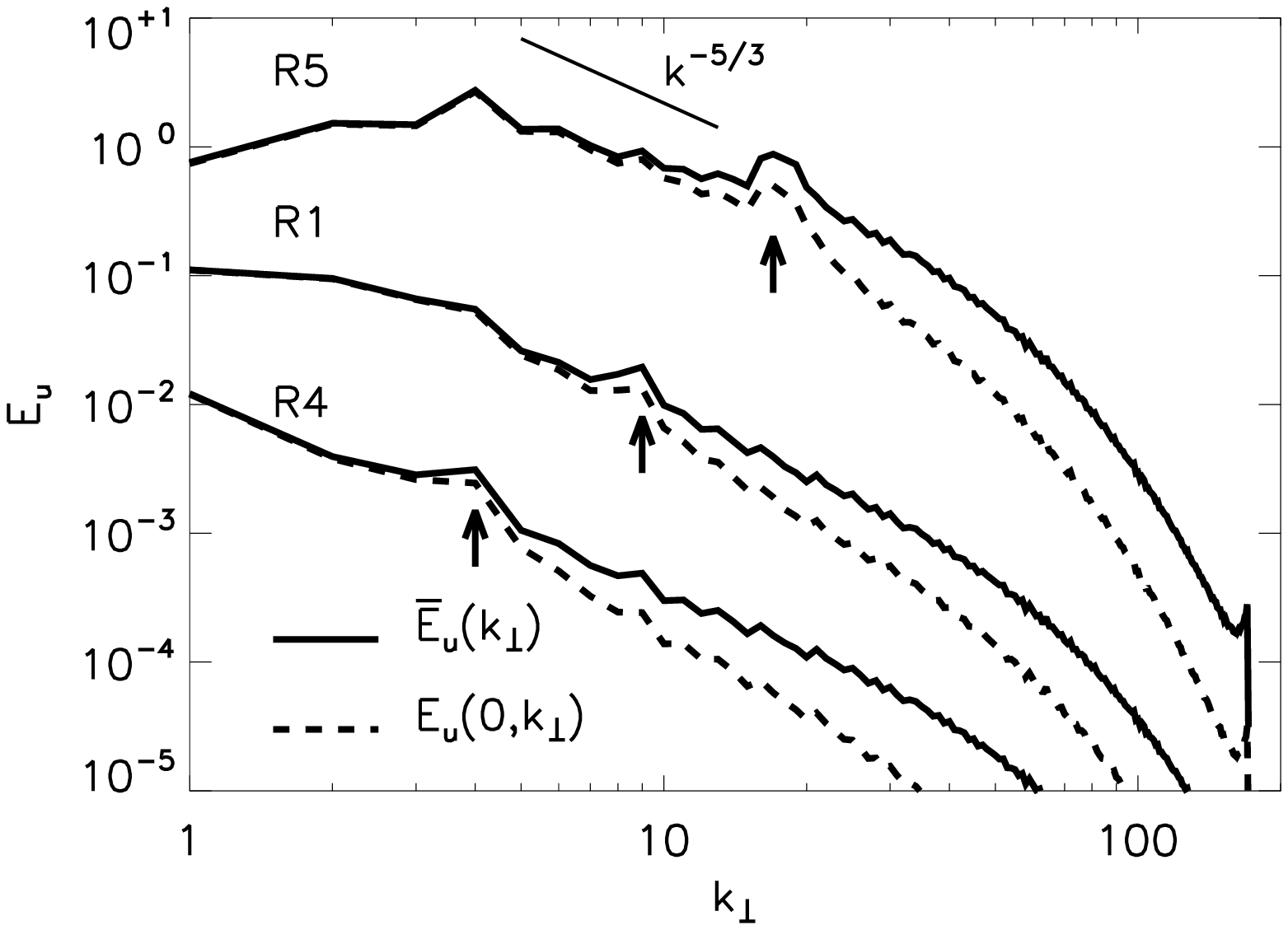}     \\                                                                               %
\includegraphics[width=8cm]{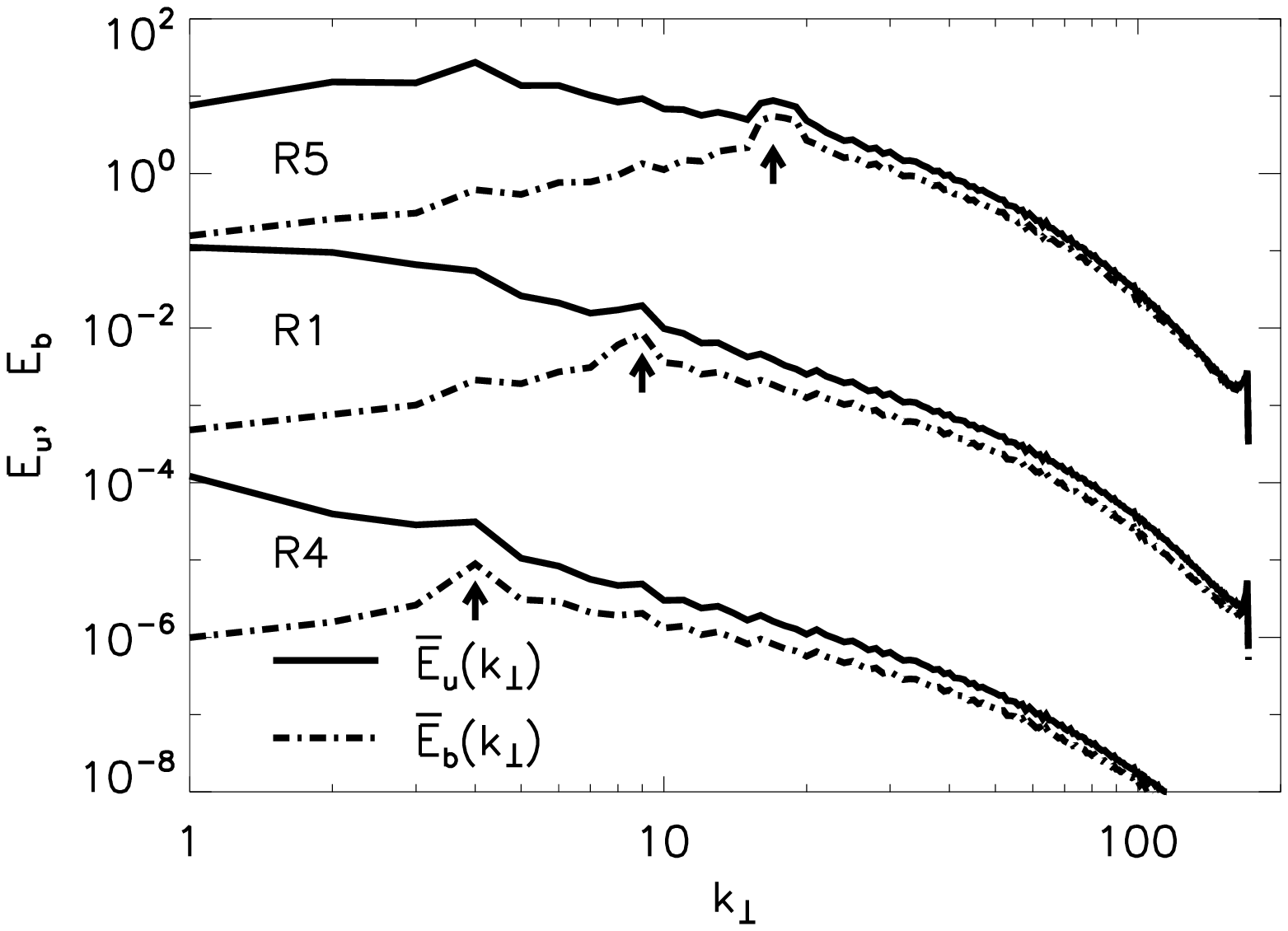}                                                                                      %
\caption{\label{C2} Top panel: The  kinetic energy spectra of $\bar{E}_u(k_\perp)$ (solid line),                              %
                                                          and      $E_u(0,k_\perp)$ (dashed line)                             %
                 for R5, (top lines), R1, (middle lines), R4, (bottom lines)                                                  %
                 Bottom panel: The kinetic energy spectra $\bar{E}_u(k_\perp)$ (solid line),                                  %
                 compared to the magnetic energy spectra $\bar{E}_b(k_\perp)$ (dashed line)                                   %
                 of the same runs and with the same order.                                                                    %
The spectra have been shifted for reasons of clarity.  The arrows indicate the location of the forcing.  }                    %
\end{figure}                                                                                                                  %
%%%%%%%%%%%%%%%%%%%%%%%%%%%%%%%%%%%%%%%%%%%%%%%%%%%%%%%%%%%%%%%%%%%%%%%%%%%%%%%%%%%%%%%%%%%%%%%%%%%%%%%%%%%%%%%%%%%%%%%%%%%%%%%

The second parameter we examine is the  forcing scale $k_fL$. This parameter is important because it controls
the number of modes that satisfy the quasi-resonance conditions (\ref{con1}),(\ref{con2}). 
The energy flux of runs R4 with $4 <k_f \le 5$ and 
R5 with $16<k_f \le 20$ are compared to the energy flux of R1 with $8<k_f\le 10$ in figure \ref{C1}.
The top panel again shows the energy flux in the perpendicular direction while the bottom panel shows the energy flux
in the parallel direction.

All flows show an inverse cascade in the perpendicular direction. R4 has an inverse cascade of the same amplitude with 
R1 while R5 that is forced in smaller scales has a weaker inverse cascade. This is somehow expected since when the 
forcing is in smaller scales the system is closer in violating condition (\ref{con2}) for 2D behavior.
Note also that R4 has a larger flux in the parallel direction.

The spectra for these runs are shown in figure \ref{C2}. 
All runs have most of the kinetic energy concentrated in the large scales
that behave like 2D-hydrodynamic flows: $\bar{E}_u(k_\perp)\simeq E_u(0,k_\perp)$ (top panel)
and $\bar{E}_u(k_\perp) \gg \bar{E}_b(k_\perp)$ (bottom panel). The scales smaller than the forcing scale on the other hand
behave like 3D-MHD flows with $\bar{E}_u(k_\perp) > E_u(0,k_\perp)$ and $\bar{E}_u(k_\perp) \simeq \bar{E}_b(k_\perp)$.

%%%%%%%%%%%%%%%%%%%%%%%%%%%%%%%%%%%%%%%%%%%%%%%%%%%%%%%%%%%%%%%%%%%%%%%%%%%%%%%%%%%%%%%%%%%%%%%%%%%%%%%%%%%%%%%%%%%%%%%%%%%%%%e
%%%%%%%%%%%%%%%%%%%%%%%%%%%%%%%%%%%%%%%%%%%%%%%%%%%%%%%%%%%%%%%%%%%%%%%%%%%%%%%%%%%%%%%%%%%%%%%%%%%%%%%%%%%%%%%%%%%%%%%%%%%%%%%
\subsection{Mechanichal and electro-motive forcing }%%%%%%%%%%%%%%%%%%%%%%%%%%%%%%%%%%%%%%%%%%%%%%%%%%%%%%%%%%%%%%%%%%%%%%%%%%%%
%%%%%%%%%%%%%%%%%%%%%%%%%%%%%%%%%%%%%%%%%%%%%%%%%%%%%%%%%%%%%%%%%%%%%%%%%%%%%%%%%%%%%%%%%%%%%%%%%%%%%%%%%%%%%%%%%%%%%%%%%%%%%%%
%%%%%%%%%%%%%%%%%%%%%%%%%%%%%%%%%%%%%%%%%%%%%%%%%%%%%%%%%%%%%%%%%%%%%%%%%%%%%%%%%%%%%%%%%%%%%%%%%%%%%%%%%%%%%%%%%%%%%%%%%%%%%%%

%%%%%%%%%%%%%%%%%%%%%%%%%%%%%%%%%%%%%%%%%%%%%%%%%%%%%%%%%%%%%%%%%%%%%%%%%%%%%%%%%%%%%%%%%%%%%%%%%%%%%%%%%%%%%%%%%%%%%%%%%%%%%%%
\begin{figure}                                                                                                                %
\includegraphics[width=8cm]{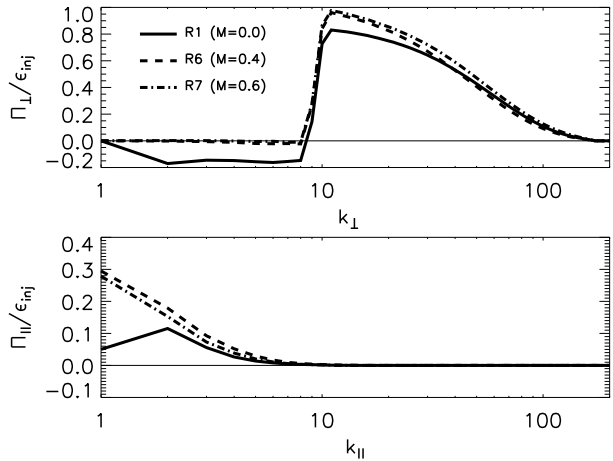}                                                                                  %
\caption{\label{D1} Top panel: The energy flux in the perpendicular direction for R1 ($M=0$,   solid line),                   %
                                                                                  R6 ($M=0.4$, dashed line),                  %
                                                                                  R7 ($M=0.6$, dashed-dot line).              %
                    Bottom panel: The energy flux in the perpendicular direction for the same runs.}                          %
\end{figure}                                                                                                                  %
%%%%%%%%%%%%%%%%%%%%%%%%%%%%%%%%%%%%%%%%%%%%%%%%%%%%%%%%%%%%%%%%%%%%%%%%%%%%%%%%%%%%%%%%%%%%%%%%%%%%%%%%%%%%%%%%%%%%%%%%%%%%%%%

%%%%%%%%%%%%%%%%%%%%%%%%%%%%%%%%%%%%%%%%%%%%%%%%%%%%%%%%%%%%%%%%%%%%%%%%%%%%%%%%%%%%%%%%%%%%%%%%%%%%%%%%%%%%%%%%%%%%%%%%%%%%%%%
\begin{figure}                                                                                                                %
\includegraphics[width=8cm]{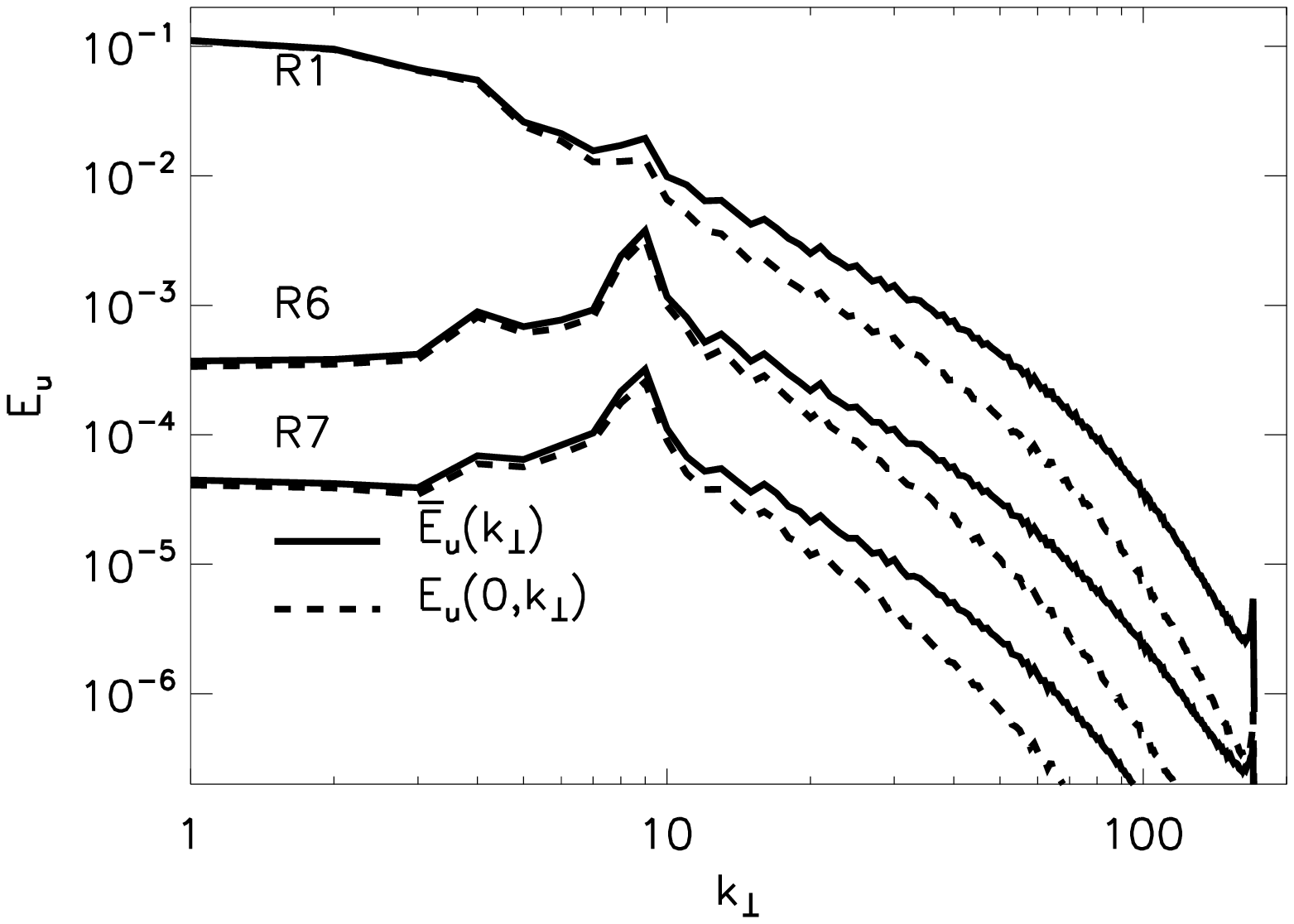}                                                                                      %
\includegraphics[width=8cm]{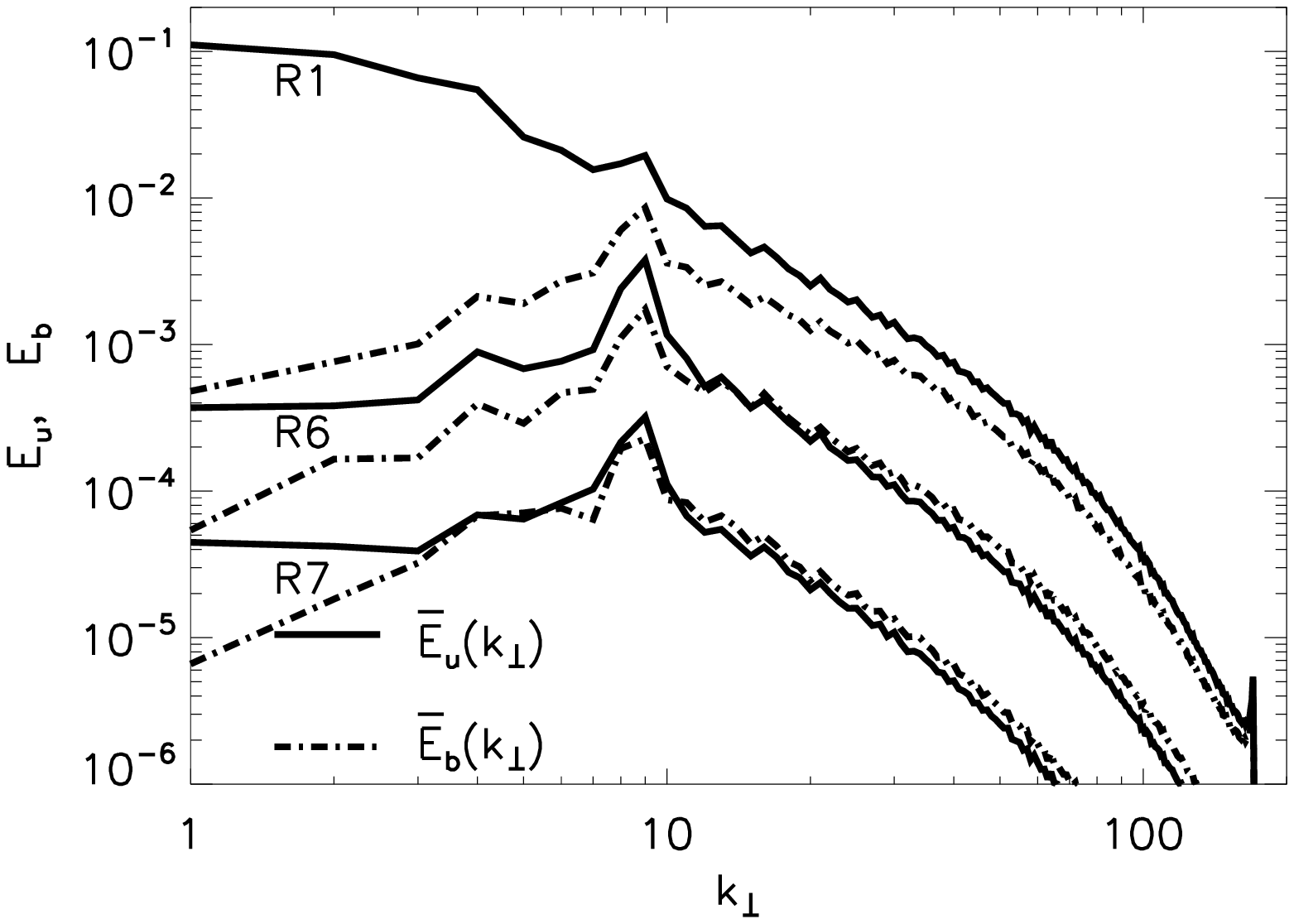}                                                                                      %
\caption{\label{D2} Top panel: The  kinetic energy spectra of $\bar{E}_u(k_\perp)$ (solid line),                              %
                                                          and      $E_u(0,k_\perp)$ (dashed line)                             %
for R1, $M=0$ (top lines), R6, $M=0.4$ (middle lines), R7 $M=0.6$ (bottom lines)                                              %
                 Bottom panel: The kinetic energy spectra $\bar{E}_u(k_\perp)$ (solid line),                                  %
                 compared to the magnetic energy spectra $\bar{E}_b(k_\perp)$ (dashed line)                                   %
                 of the same runs and with the same order.                                                                    %
The spectra have been shifted for reasons of clarity.     }                                                                   %
\end{figure}                                                                                                                  %
%%%%%%%%%%%%%%%%%%%%%%%%%%%%%%%%%%%%%%%%%%%%%%%%%%%%%%%%%%%%%%%%%%%%%%%%%%%%%%%%%%%%%%%%%%%%%%%%%%%%%%%%%%%%%%%%%%%%%%%%%%%%%%%

%%%%%%%%%%%%%%%%%%%%%%%%%%%%%%%%%%%%%%%%%%%%%%%%%%%%%%%%%%%%%%%%%%%%%%%%%%%%%%%%%%%%%%%%%%%%%%%%%%%%%%%%%%%%%%%%%%%%%%%%%%%%%%%
\begin{figure}                                                                                                                %
\includegraphics[width=8cm]{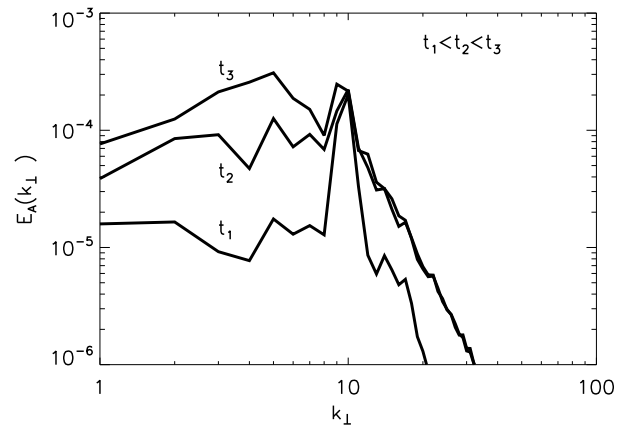}                                                                                       %
\caption{\label{D3} The vector potential spectrum $\bar{E}_{_A}$ for three different times.              }                    %
\end{figure}                                                                                                                  % 
%%%%%%%%%%%%%%%%%%%%%%%%%%%%%%%%%%%%%%%%%%%%%%%%%%%%%%%%%%%%%%%%%%%%%%%%%%%%%%%%%%%%%%%%%%%%%%%%%%%%%%%%%%%%%%%%%%%%%%%%%%%%%%% 

The inverse cascade observed in some of the discussed runs is a property of 2D hydrodynamic turbulence that is not
present in 2D-MHD turbulence.
The reason it appears in the previous runs is that the amplitude
of the magnetic field fluctuations in the large scales remains weak. 
This effect could possibly be destroyed by a large scale dynamo at larger magnetic Reynolds numbers.
Leaving this possibility open the existence of the inverse cascade is investigated when magnetic field fluctuations 
are amplified by an electro-motive force. 
This is examined in runs R6 and R7 where both the mechanical and the electro-motive force are present.
  
Figure \ref{D1} shows the energy flux for runs R6 with $M=0.4$ and R7 with $M=0.6$ compared with R1 for which $M=0$.
The introduction of the electro-motive force ($M\ne0$) destroys the inverse cascade in the perpendicular 
direction (top panel), while little change is observed in the parallel direction (bottom panel). This change indicates
that the system transitions from an hydrodynamic 2D state to a forward cascading MHD state.

This is further confirmed by looking at the energy spectra in figure \ref{D2}. The top panel compares again the
kinetic energy spectra $\bar{E_u}$ and $E_u(0,k_\perp)$. 
The excess of kinetic energy that is present in the large scales for run R1, is absent in runs R6 and R7, verifying further
the absence of the inverse cascade in the presence of an electro-motive force. Note that in all runs 
the large scales are still two-dimensional  $\bar{E_u} \simeq E(0,k_\perp)$ (top panel) but the condition 
$\bar{E}_u (k_\perp) \gg \bar{E}_b(k_\perp)$ is true only for R1 (bottom panel).
This indicates that the absence of the inverse cascades for runs R6 and R7 is not because
the flow stops behaving like a 2D flow, but rather because it starts behaving like a 2D-MHD flow.

In 2D-MHD flows however there is an inverse cascade of the square of the vector potential that is a conserved quantity.
If the flow in the runs R6 and R7 behave like a 2D MHD flow in the large scales such a cascade should be observed.
However a flux for the squared vector potential in three dimensions can not be uniquely defined since it is not a conserved quantity. Nonetheless, we plot the vector potential spectra for three different times from run R7 in figure \ref{D3}.
The vector potential ${\bf a}$ is defined so that ${\bf b = \nabla \times a}$ and ${\bf \nabla \cdot a=}0$.
Its spectrum is then defined as
\beq
\bar{E}_{_A}(k_\perp)= \frac{1}{2} \sum {\bf |\hat{a}_k}|^2 
\eeq 
where the sum is restricted in the wavenumbers $k_\perp \le \sqrt{k_x^2+k_y^2} < k_\perp+1$ and 
${\bf \hat{a}_k }$ is the Fourier transform of ${\bf a}$.
It can be seen that as time progresses the vector potential is increasing in the large scales.
It is noted that a quasi-conservation of the square of the vector potential has been observed in
\cite{Servidio2005,Dmitruk2011} for three-dimensional ideal reduced-MHD.

%%%%%%%%%%%%%%%%%%%%%%%%%%%%%%%%%%%%%%%%%%%%%%%%%%%%%%%%%%%%%%%%%%%%%%%%%%%%%%%%%%%%%%%%%%%%%%%%%%%%%%%%%%%%%%%%%%%%%%%%%%%%%%%
%%%%%%%%%%%%%%%%%%%%%%%%%%%%%%%%%%%%%%%%%%%%%%%%%%%%%%%%%%%%%%%%%%%%%%%%%%%%%%%%%%%%%%%%%%%%%%%%%%%%%%%%%%%%%%%%%%%%%%%%%%%%%%%
\subsection{Isotropic and anisotropic forcing}%%%%%%%%%%%%%%%%%%%%%%%%%%%%%%%%%%%%%%%%%%%%%%%%%%%%%%%%%%%%%%%%%%%%%%%%%%%%%%%%%
%%%%%%%%%%%%%%%%%%%%%%%%%%%%%%%%%%%%%%%%%%%%%%%%%%%%%%%%%%%%%%%%%%%%%%%%%%%%%%%%%%%%%%%%%%%%%%%%%%%%%%%%%%%%%%%%%%%%%%%%%%%%%%%
%%%%%%%%%%%%%%%%%%%%%%%%%%%%%%%%%%%%%%%%%%%%%%%%%%%%%%%%%%%%%%%%%%%%%%%%%%%%%%%%%%%%%%%%%%%%%%%%%%%%%%%%%%%%%%%%%%%%%%%%%%%%%%%
\label{sect}
%%%%%%%%%%%%%%%%%%%%%%%%%%%%%%%%%%%%%%%%%%%%%%%%%%%%%%%%%%%%%%%%%%%%%%%%%%%%%%%%%%%%%%%%%%%%%%%%%%%%%%%%%%%%%%%%%%%%%%%%%%%%%%%
\begin{figure}                                                                                                                %
\includegraphics[width=8cm]{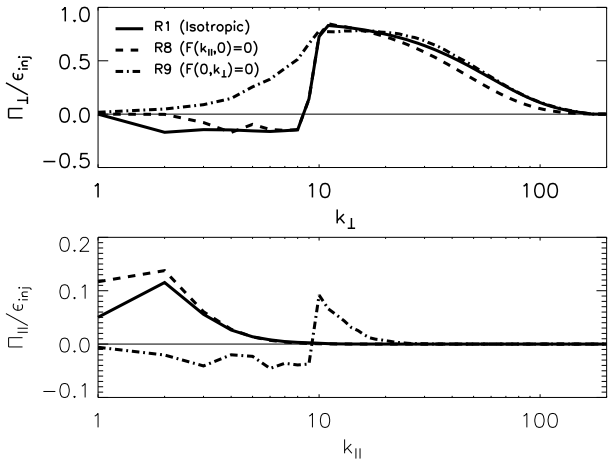}                                                                                       %
\caption{\label{E1} Top panel: The energy flux in the perpendicular direction for R1 (solid line),                            %
                                                                                  R8 (dashed line),                           %
                                                                                  R9 (dashed-dot line).                       %
                    Bottom panel: The energy flux in the perpendicular direction for the same runs.}                          %
\end{figure}                                                                                                                  %  
%%%%%%%%%%%%%%%%%%%%%%%%%%%%%%%%%%%%%%%%%%%%%%%%%%%%%%%%%%%%%%%%%%%%%%%%%%%%%%%%%%%%%%%%%%%%%%%%%%%%%%%%%%%%%%%%%%%%%%%%%%%%%%% 

%%%%%%%%%%%%%%%%%%%%%%%%%%%%%%%%%%%%%%%%%%%%%%%%%%%%%%%%%%%%%%%%%%%%%%%%%%%%%%%%%%%%%%%%%%%%%%%%%%%%%%%%%%%%%%%%%%%%%%%%%%%%%%%
\begin{figure}                                                                                                                %
\includegraphics[width=8cm]{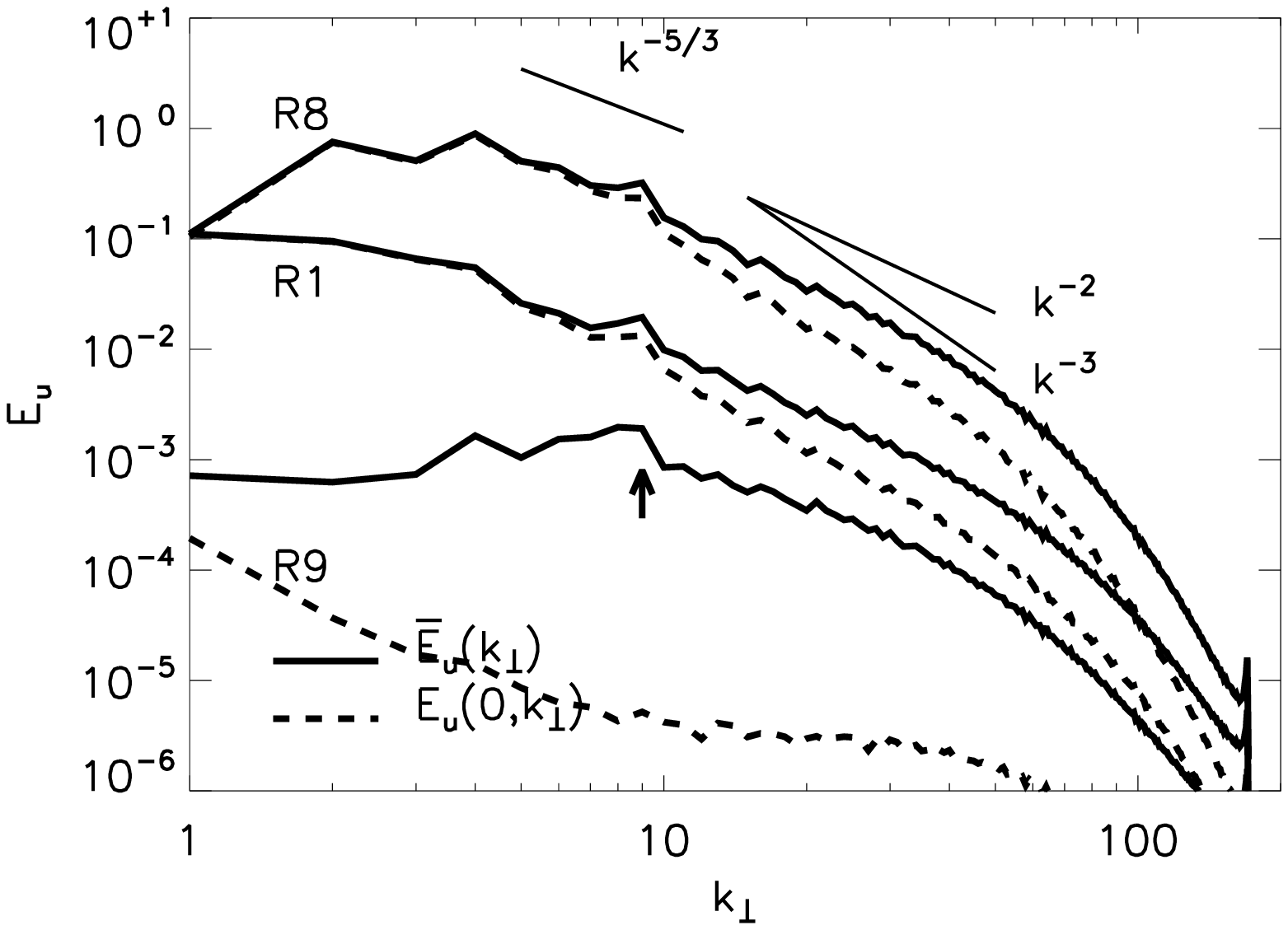}                                                                             %
\includegraphics[width=8cm]{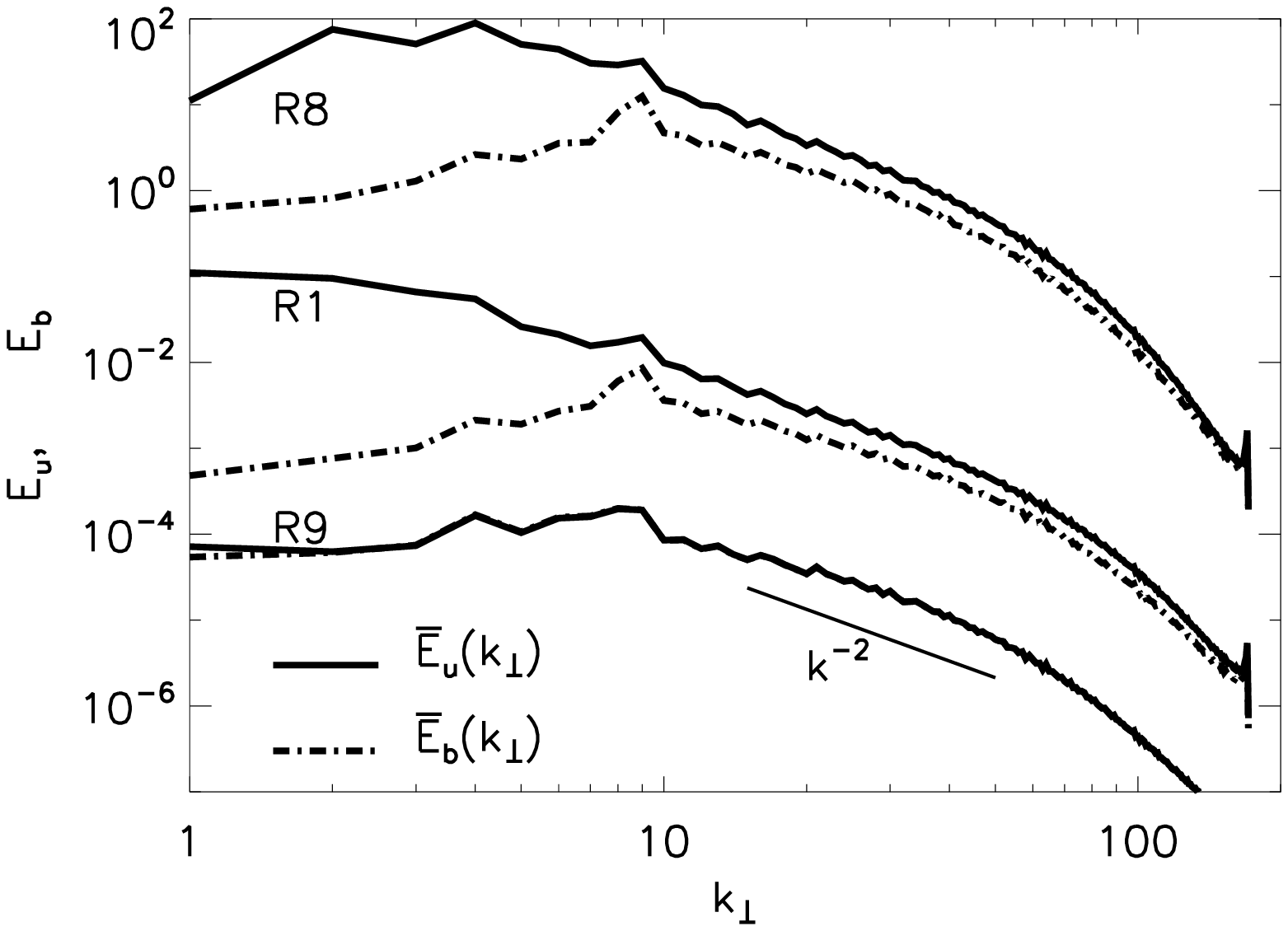}                                                                             %
\caption{\label{E2} Top panel: The  kinetic energy spectra of $\bar{E}_u(k_\perp)$ (solid line),                              %
                                                          and      $E_u(0,k_\perp)$ (dashed line)                             %
for R8, (top lines), R1, (middle lines), R9 (bottom lines)                                                                    %
                 Bottom panel: The kinetic energy spectra $\bar{E}_u(k_\perp)$ (solid line),                                  %
                 compared to the magnetic energy spectra $\bar{E}_b(k_\perp)$ (dashed line)                                   %
                 of the same runs and with the same order.                                                                    %
The spectra have been shifted for reasons of clarity.     }                                                                   %
\end{figure}                                                                                                                  %
%%%%%%%%%%%%%%%%%%%%%%%%%%%%%%%%%%%%%%%%%%%%%%%%%%%%%%%%%%%%%%%%%%%%%%%%%%%%%%%%%%%%%%%%%%%%%%%%%%%%%%%%%%%%%%%%%%%%%%%%%%%%%%%

The last parameter that we vary is the isotropy of the forcing. Unlike the previously examined runs for which
all Fourier modes within a spherical shell are uniformly forced in runs R8 and R9 the modes $k_\|=0$ 
and $k_\perp=0$ respectably are forced preferentially. In particular the amplitude of a Fourier mode ${\bf F_k}$
of the forcing inside the chosen spherical shell was proportional to:
\beq
\mathrm{R8}:\,\, | {\bf F_k} | \propto \frac{ k_x^2 + k_y^2 }{k^2}, \quad
\mathrm{R9}:\,\, |{\bf F_k} |  \propto \frac{ k_z^2 }{k^2},
\eeq
where $k^2=k_x^2+k_y^2+k_z^2$.
Thus in run R8 the 2D modes ($k_z=0$) are forced preferentially while 
     in run R9 these 2D modes are not forced at all and most forcing is at the high $k_z$ modes.

The energy flux for these runs is shown in figure \ref{E1}. The energy flux of run R1 and R8 are very close for 
both directions.
The reason for the slightly smaller flux of R8 at the largest scales is because R8 was evolved for a shorter time than R1. 
R9 however has no inverse cascade in the perpendicular direction (top panel). 
This is not surprising since for this run the 2D modes are not forced at all.
In the parallel direction there is also a drastic change. 
R9 shows an inverse cascade from the large $k_z$ wavenumbers to the 2D $k_z=0$ modes. 
%This result is somehow unexpected since it has not been predicted by any of the current theories.
It is noted that the flux towards the large parallel  scales was strongly fluctuating taking both positive and negative values.
Only after averaging several files the result shown in figure \ref{E1} was obtained.

The spectra for these runs are shown in figure \ref{E2}. Again not a lot of difference can be seen between run R1 and R8.
Both are close to two-dimensional in the large scales and three dimensional in the small scales (top panel), 
and both have weak magnetic energy in the large scales but are close to equipartition in the small scales (bottom panel).
R9 on the other hand is three dimensional for all scales and kinetic and magnetic energy are almost identically equal
at all scales. This last remark indicates that Alfven-waves (for which ${\bf u =\pm b}$) dominate the turbulence.

%%%%%%%%%%%%%%%%%%%%%%%%%%%%%%%%%%%%%%%%%%%%%%%%%%%%%%%%%%%%%%%%%%%%%%%%%%%%%%%%%%%%%%%%%%%%%%%%%%%%%%%%%%%%%%%%%%%%%%%%%%%%%%%
%%%%%%%%%%%%%%%%%%%%%%%%%%%%%%%%%%%%%%%%%%%%%%%%%%%%%%%%%%%%%%%%%%%%%%%%%%%%%%%%%%%%%%%%%%%%%%%%%%%%%%%%%%%%%%%%%%%%%%%%%%%%%%%
%%%%%%%%%%%%%%%%%%%%%%%%%%%%%%%%%%%%%%%%%%%%%%%%%%%%%%%%%%%%%%%%%%%%%%%%%%%%%%%%%%%%%%%%%%%%%%%%%%%%%%%%%%%%%%%%%%%%%%%%%%%%%%%
%%%%%%%%%%%%%%%%%%%%%%%%%%%%%%%%%%%%%%%%%%%%%%%%%%%%%%%%%%%%%%%%%%%%%%%%%%%%%%%%%%%%%%%%%%%%%%%%%%%%%%%%%%%%%%%%%%%%%%%%%%%%%%%
%%%%%%%%%%%%%%%%%%%%%%%%%%%%%%%%%%%%%%%%%%%%%%%%%%%%%%%%%%%%%%%%%%%%%%%%%%%%%%%%%%%%%%%%%%%%%%%%%%%%%%%%%%%%%%%%%%%%%%%%%%%%%%%
%%%%%%%%%%%%%%%%%%%%%%%%%%%%%%%%%%%%%%%%%%%%%%%%%%%%%%%%%%%%%%%%%%%%%%%%%%%%%%%%%%%%%%%%%%%%%%%%%%%%%%%%%%%%%%%%%%%%%%%%%%%%%%%
\section{Summary and Discussion}
%%%%%%%%%%%%%%%%%%%%%%%%%%%%%%%%%%%%%%%%%%%%%%%%%%%%%%%%%%%%%%%%%%%%%%%%%%%%%%%%%%%%%%%%%%%%%%%%%%%%%%%%%%%%%%%%%%%%%%%%%%%%%%%
%%%%%%%%%%%%%%%%%%%%%%%%%%%%%%%%%%%%%%%%%%%%%%%%%%%%%%%%%%%%%%%%%%%%%%%%%%%%%%%%%%%%%%%%%%%%%%%%%%%%%%%%%%%%%%%%%%%%%%%%%%%%%%%
%%%%%%%%%%%%%%%%%%%%%%%%%%%%%%%%%%%%%%%%%%%%%%%%%%%%%%%%%%%%%%%%%%%%%%%%%%%%%%%%%%%%%%%%%%%%%%%%%%%%%%%%%%%%%%%%%%%%%%%%%%%%%%%
%%%%%%%%%%%%%%%%%%%%%%%%%%%%%%%%%%%%%%%%%%%%%%%%%%%%%%%%%%%%%%%%%%%%%%%%%%%%%%%%%%%%%%%%%%%%%%%%%%%%%%%%%%%%%%%%%%%%%%%%%%%%%%%
%%%%%%%%%%%%%%%%%%%%%%%%%%%%%%%%%%%%%%%%%%%%%%%%%%%%%%%%%%%%%%%%%%%%%%%%%%%%%%%%%%%%%%%%%%%%%%%%%%%%%%%%%%%%%%%%%%%%%%%%%%%%%%%
%%%%%%%%%%%%%%%%%%%%%%%%%%%%%%%%%%%%%%%%%%%%%%%%%%%%%%%%%%%%%%%%%%%%%%%%%%%%%%%%%%%%%%%%%%%%%%%%%%%%%%%%%%%%%%%%%%%%%%%%%%%%%%%

In this work we have shown that under certain conditions an MHD-flow in the presence of a strong magnetic field can behave
like a two dimensional flow in the large scales while like three dimensional
(possibly weak turbulence) in the small scales, much like strongly rotating fluids.
In the large scales it was found that the magnetic fluctuations were suppressed
and a 2D inverse cascade of energy developed with energy accumulating in the large scales. 

This inverse cascade however is sensitive to various parameters.
If the uniform magnetic field amplitude is decreased sufficiently,  
or if the domain size is increased (or equivalently the forcing scale is decreased)
so that the  condition (\ref{con2}) for two-dimensionalization  no longer holds 
the flow recovers its 3D behavior and cascades the energy forward. 
Furthermore, in the case that an electro-motive force is introduced  although the flow remains  
2D in the large-scales the magnetic fluctuations are no longer suppressed and the flow behaves like a 2D-MHD fluid with a 
direct energy cascade. At the same time an indication of an inverse cascade of the squared vector potential was observed.
Finally, absence of an inverse cascade in the perpendicular direction was also observed when the system was forced only in 
the large $k_\|$ modes. In this case an inverse cascade in the direction parallel to the magnetic field was observed.

All the simulations presented here that exhibited an inverse cascade, were stopped before 
the largest scale of the system was reached due the computational cost. If the runs were continued 
for a longer time as the energy and the perpendicular wave 
number are increased it is possible that a point will be reached that the strength of the magnetic field will not be sufficient to
stop three dimensional instabilities from breaking the two dimensional constrain.
Such a transition point is expected to appear when the eddy turn over frequency $u_kk_\perp$ becomes smaller than the Alfven
frequency $B_0k_\|$.
In the inertia range of the inverse 2D cascade where the $k_\perp^{-5/3}$ scaling is expected 
the eddy turn over frequency is decreasing as larger scales are reached. 
On the contrary the smallest Alfven frequency $B_0k_\|$ remains independent of $k_\perp$. 
Thus the ratio $u_kk_\perp/B_0k_\|$ will decrease as the cascade proceeds and it
is not expected that such a transition point will exist in the inertial range, instead 
as the cascade proceeds the flow will come closer to a 2D flow.
However since there is no large-scale damping mechanism to dissipate the energy when the largest scale of the system is reached 
the energy would pile up in this scale. In this case the eddy turn over frequency will increase and eventually the criterion
(\ref{con2}) for two-dimensionality will be violated. Then energy could possibly return to the small scales as weak or strong 
turbulence. Similar scenario for the fate of the inverse cascade of rotating turbulence has been proposed in \cite{Nazarenko2011}. 
Such scenarios however need to be verified by numerical simulations and experiments. 

\begin{acknowledgments}
Computations were carried out on the CEMAG computing center on the 
CINES computing center, and their support is greatly acknowledged.
The author would also like to thank the participants of the participants
of the ``Dynamics and turbulent transport in  plasmas and conducting fluids"
workshop in Les Houches, France March 2011 for their useful discussions and
suggestions.
\end{acknowledgments}

%%%%%%%%%%%%%%%%%%%%%%%%%%%%%%%%%%%%%%%%%%%%%%%%%%%%%%%%%%%%%%%%%%%
%%%%%%%%%%%%%%%%%%%%%%%%%%%%%%%%%%%%%%%%%%%%%%%%%%%%%%%%%%%%%%%%%%%

%\begin{thebibliography}{99}

\bibliography{MHDturb}

%\end{thebibliography}
\end{document}